\documentclass[reprint,amsmath,amssymb,aps,prx,superscriptaddress]{revtex4-2}
\usepackage{graphicx}
\usepackage{dcolumn}
\usepackage{bm}
\usepackage[colorlinks,linkcolor=blue,urlcolor=blue,anchorcolor=black,citecolor=blue]{hyperref}
\usepackage{url}
\usepackage{subfigure}
\usepackage{color}

\begin{document}

\title{Self-supervised Representations and Node Embedding Graph Neural Networks for Accurate and Multi-scale Analysis of Materials}

\author{Jian-Gang Kong}
\affiliation{School of Science, Chongqing University of Posts and Telecommunications, Chongqing 400065, China}%
\author{Ke-Lin Zhao}
\affiliation{School of Science, Chongqing University of Posts and Telecommunications, Chongqing 400065, China}%
\author{Jian Li}
\affiliation{School of Science, Chongqing University of Posts and Telecommunications, Chongqing 400065, China}%
\affiliation{Institute for Advanced Sciences, Chongqing University of Posts and Telecommunications, Chongqing 400065, China}
\affiliation{Southwest Center for Theoretical Physics, Chongqing University, Chongqing 401331, China}
\author{Qing-Xu Li}
\affiliation{School of Science, Chongqing University of Posts and Telecommunications, Chongqing 400065, China}%
\affiliation{Institute for Advanced Sciences, Chongqing University of Posts and Telecommunications, Chongqing 400065, China}%
\author{Yu Liu}
\affiliation{Inspur Electronic Information Industry Co., Ltd, Beijing 100085, China}
\author{Rui Zhang}
\email{zhangrui727@huawei.com}
\affiliation{HiSilicon Technologies Co., Ltd. Shenzhen, China}
\author{Jia-Ji Zhu}
\email{zhujj@cqupt.edu.cn}
\affiliation{School of Science, Chongqing University of Posts and Telecommunications, Chongqing 400065, China}%
\affiliation{Institute for Advanced Sciences, Chongqing University of Posts and Telecommunications, Chongqing 400065, China}
\affiliation{Southwest Center for Theoretical Physics, Chongqing University, Chongqing 401331, China}
\author{Kai Chang}
\email{kchang@zju.edu.cn}
\affiliation{School of Physics, Zhejiang University, Hangzhou 310027, China}

\begin{abstract}
Supervised machine learning algorithms, such as graph neural networks (GNN), have successfully predicted material properties. However, the superior performance of GNN usually relies on end-to-end learning on large material datasets, which may lose the physical insight of multi-scale information about materials. And the process of labeling data consumes many resources and inevitably introduces errors, which constrains the accuracy of prediction. We propose to train the GNN model by self-supervised learning on the node and edge information of the crystal graph. Compared with the popular manually constructed material descriptors, the self-supervised atomic representation can reach better prediction performance on material properties. Furthermore, it may provide physical insights by tuning the range information. Applying the self-supervised atomic representation on the magnetic moment datasets, we show how they can extract rules and information from the magnetic materials. To incorporate rich physical information into the GNN model, we develop the node embedding graph neural networks (NEGNN) framework and show significant improvements in the prediction performance. The self-supervised material representation and the NEGNN framework may investigate in-depth information from materials and can be applied to small datasets with increased prediction accuracy.
\end{abstract}

\maketitle

\section{Introduction}
The traditional way of material design strongly relies on the experience and intuition of experts. Trial-and-error experiments are performed to synthesize new materials, usually requiring a long period and high cost. The traditional computational methods are based on domain knowledge, including density functional theory, molecular dynamics, and density matrix renormalization group. A new methodology, the machine learning method, has emerged in material science and  has been successfully applied to predict various properties of materials\cite{PilaniaG} and the inverse design of materials aiming at specific properties\cite{NohJ}.

The data representation of materials, the so-called material descriptor, is one of the core elements for the application of machine learning in materials science\cite{GhiringhelliL}. Manually constructed material descriptors can be classified into two categories---local atomic descriptors and global descriptors\cite{CeriottiM, MusilF, FMusil, LangerM}. The local atomic descriptor\cite{BehlerJ, DeS}, e.g., orbital field matrix (OFM)\cite{PhamT}, describes an atom and its surrounding environment, which can be directly used as the input of machine learning algorithm to study physical properties at the atomic level\cite{UnkeO, PhamT,rkle}. The global descriptor describes bondings or interactions between atoms, such as Coulomb matrix (CM)\cite{RuppM} and sine matrix (SM)\cite{FaberF}, which are suited for the global properties of the system.

The deep learning method\cite{LeCunY}, a rising star of the machine learning family, has also achieved remarkable performance in materials science\cite{ChoudharyK1} with its huge number of trainable parameters and high non-linearity. Moreover, compared to manually constructed material descriptors, deep neural networks can be trained to extract high-level material representations relevant to the target material properties in an automatic way\cite{SaucedaH, ChenC}. Recently, Xie et al. \cite{XieT} developed a crystal graph convolutional neural networks (CGCNN) framework to encode crystalline materials by using graph-structured data. The CGCNN has achieved excellent performance in predicting several material properties, for instance, formation energy and band gap. 

Despite recent interesting developments in graph machine learning methods for molecular systems\cite{Gasteiger2021,Batatia2022,Musaelian2023,Batzner2022}, there is still room for improvements in applying GNN to materials science. First, the power of GNN relies on large datasets typically containing more than 10,000 materials\cite{DunnA}, whereas experimental or high-fidelity computational datasets with more physical information are usually small. The GNN on small datasets often overfits and significantly complicates the training process. As a result, GNN loses its advantage over standard machine learning algorithms combined with manually constructed descriptors if the dataset is not large enough\cite{FungV}. Second, the popular graph machine learning algorithms in material science are primarily supervised learning that relies on data with labels, so they cannot fully utilize the large amount of material data that has not been labeled through calculations or experiments. On the other hand, labeling data requires high experimental or computational costs in material science, and the labeling process inevitably causes errors simultaneously. Therefore, it is crucial to reduce the reliance on labeled data.

The unsupervised pre-training strategy serves as an effective solution to reduce the reliance on labeled data, which may substantially enhance the prediction ability of the trained model. For example, the recently developed ChatGPT is a powerful and incredibly insightful conversational AI\cite{chatgpt}. As a variation of the GPT series\cite{brown2020language}, the generalization ability of ChatGPT is undoubtedly rooted in the generative pre-training (GPT) procedure on the unprecedentedly massive dataset. While unsupervised or generative pre-training has led to tremendous success in natural language processing\cite{DevlinJ} and computer vision\cite{HeK}, this methodology still needs to be widely taken in material science. Self-supervised learning on graphs is a type of unsupervised learning which constructs learning tasks by utilizing the node information, edge information of the graph itself\cite{HuW}, or augmentations of the graph\cite{YouY}. The self-supervised pre-training strategy may make it a cutting-edge research area in the graph machine learning community\cite{WuL} and possess the potential to train GNN models for materials. On the other hand, we always strive to open the black box of machine learning models to gain more insights, which is also one of the central goals of interpretable machine learning\cite{Murdoch2019}. This interpretability is particularly important in the field of machine learning for material science, as understanding the underlying physical mechanisms of target material properties can help us design materials with better performance\cite{Oviedo2022}. Therefore, constructing a  machine learning model for material science that is both accurate and interpretable is of significant importance.

Here we propose a self-supervised learning strategy on crystal graphs for a general multi-scale atomic representation using a computational material dataset. This strategy enables the GNN to effectively capture the rules of elements and the local structure in materials by recovering the randomly masked atomic properties and the distance between atoms. The trained GNN generates fixed-length, low-dimensional vector representations for atoms in materials, storing physical knowledge transferred from the GNN model. By concatenating the atomic representations of different GNN layers, the strategy can effectively alleviate the over-smoothing problem caused by the deep GNN. It can also fuse information within different spatial ranges, which helps better capture the complex interactions in materials and gain physical insights. Compared with the manually constructed descriptors, the self-supervised learning strategy has several benefits: (1) The length of the vector representation from the manually constructed descriptors increases rapidly with the total number of elements in the dataset or the maximum number of atoms in the unit cell, and may hinder its applications to more diverse material datasets. Nevertheless, the length of the self-supervised atomic representation is fixed. (2) Local atomic descriptors only encode the nearest neighbor information, while our proposed multi-scale self-supervised atomic representation may involve larger-scale information relevant to specific properties. (3) Manually constructed descriptors usually incorporate domain-specific knowledge, while self-supervised learning can extract more general information from materials. Then we generate a dataset of experimental magnetic moments as an illustrative example. The self-supervised atomic representations can be averaged to study various material properties with higher accuracy and lower computational cost than usual manually constructed descriptors. They can also further boost the performance of the GNN model by combining it with the graph structure of materials, which leads us to a node embedding graph neural network (NEGNN) framework.

\begin{figure*}[htbp]
	\centering
	\includegraphics[height=6.0cm,width=18.4cm]{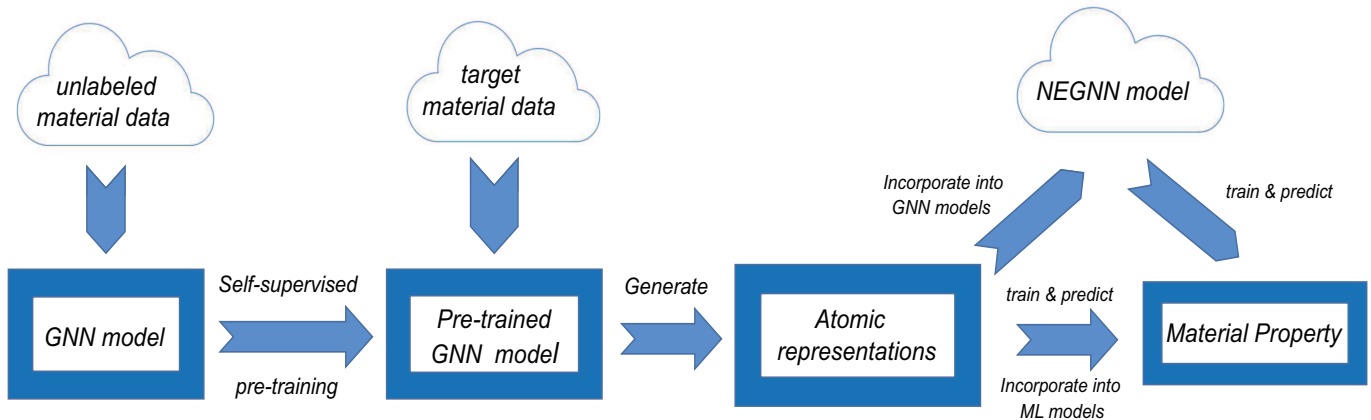}
	\caption{The high-level pipeline of the proposed NEGNN framework.}
	\label{pipeline}
\end{figure*}

The paper is organized as follows: In Sec.II, we propose the self-supervised learning strategy for the crystal graph neural network. In Sec.III-V, we generate self-supervised atomic embeddings(NEs) from the self-supervised pre-trained graph neural networks. We show its advantages by the prediction performance on magnetic moments, and visualizing the low dimensional plots. Sec.VI demonstrates the benefits of multiscale representations. In Sec.VII, we test the self-supervised graph embeddings (GEs) for material properties and develop the NEGNN framework. Finally, we summarize our framework and make comparisons of the NEGNN with some other popular GNN models in Sec.VIII. Figure \ref{pipeline} illustrates the high-level pipeline of our work.

\section{Self-supervised learning on crystal graphs} 

We propose a new strategy of performing self-supervised learning on crystal graphs using a computational material dataset from the Materials Project\cite{JainA}. The dataset size is chosen to be 60,000, which is not large compared with a typical self-supervised learning task but large enough to illustrate the effectiveness of our strategy. The types of self-supervised learning tasks on graphs can be classified into three categories: generative, contrastive, and predictive\cite{WuL}. Our strategy are mainly based on the predictive tasks for simplicity.

\begin{figure}[htbp]
	\centering
	\subfigure[]{
		\begin{minipage}[t]{1.0\linewidth}
			\centering
			\includegraphics[height=4.0cm,width=9.2cm]{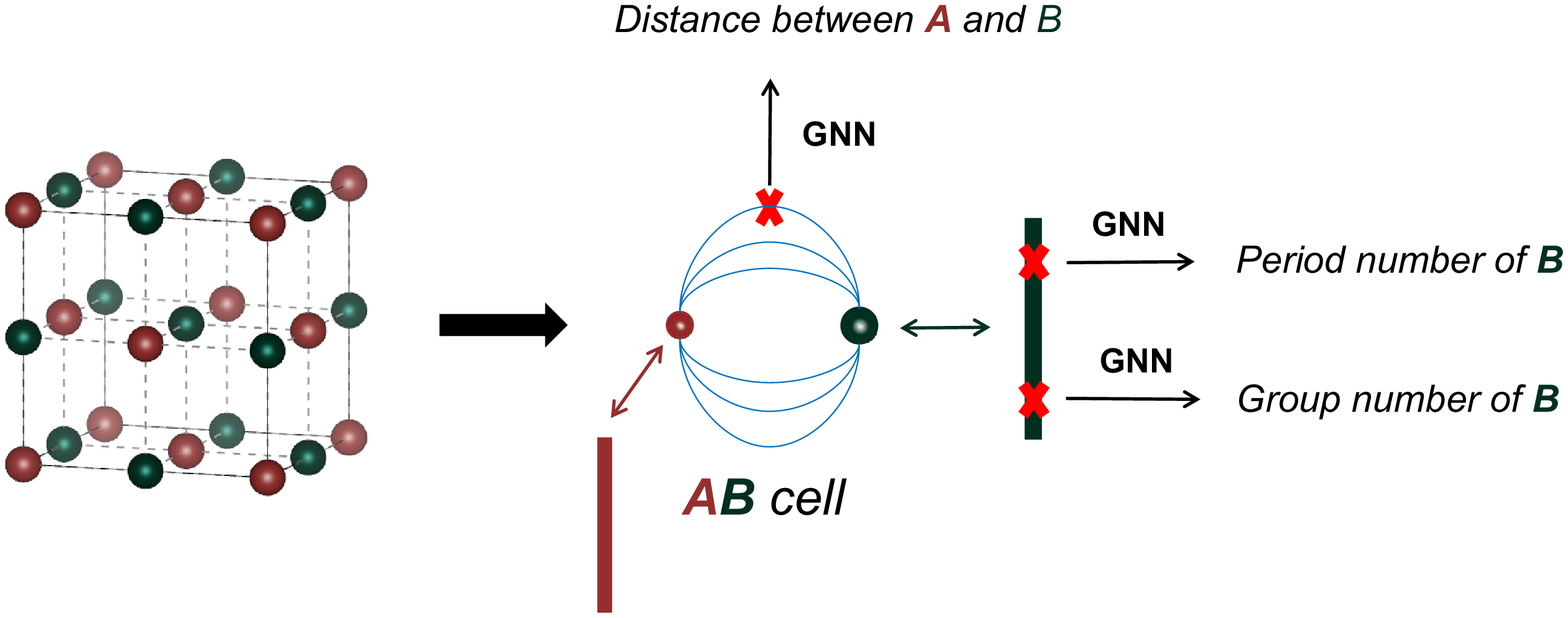}
		\end{minipage}
		\label{SL}
	}
	\hspace{.10in}
	\subfigure[]{
		\begin{minipage}[t]{1.0\linewidth}
			\centering
			\includegraphics[height=4.0cm,width=9.0cm]{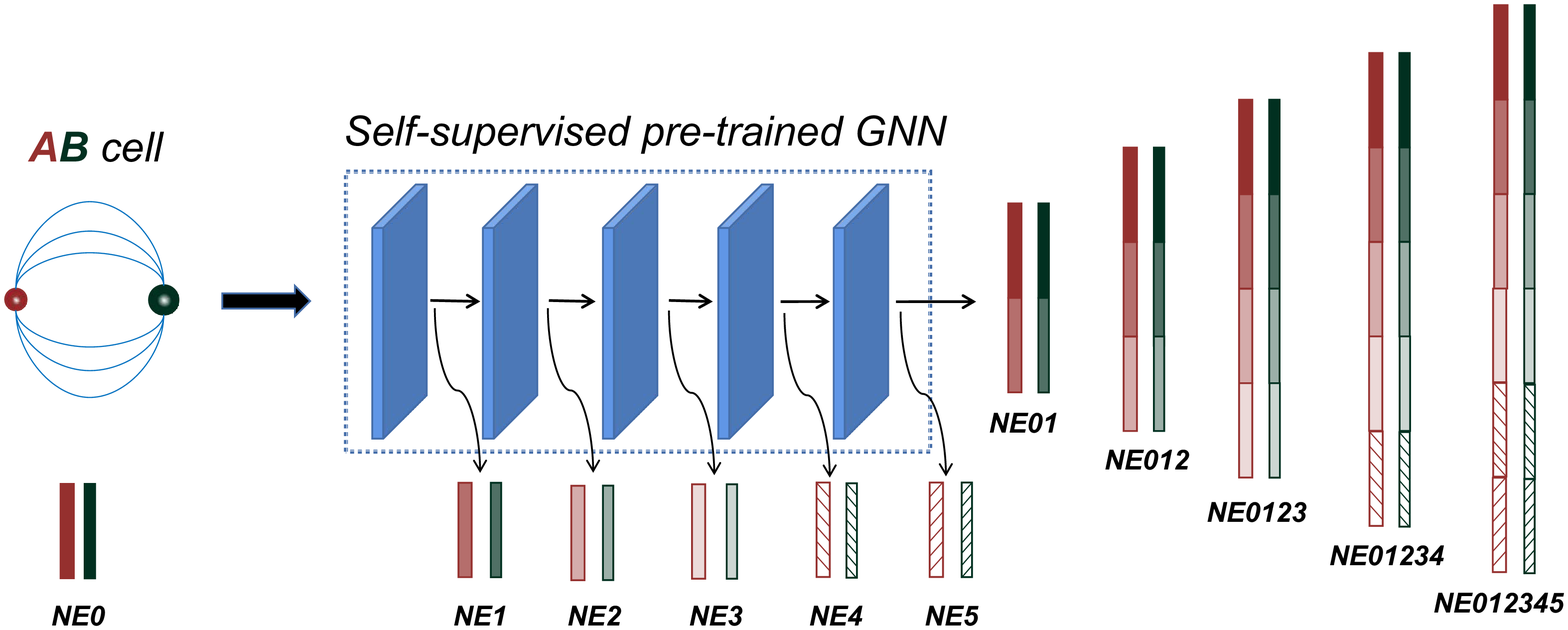}
			\label{Atomic_Vector}
		\end{minipage}
		\label{GE}	
	}
	\caption{(a) Predictive self-supervised training on crystal graphs (AB primitive cell as an illustrative example), where the dark green (dark red) balls  represent B (A) atoms, the dark green (dark red) bar is the node vector corresponding to B (A) atom, the red crosses denote randomly masked node or edge information during training. A GNN is trained to recover the masked information based on surrounding crystal environment. (b) 64-dimensional atomic vectors generated from the self-supervised pre-trained GNN given a crystal graph as input (AB primitive cell as an illustrative example), the dark red (dark green) bars at the bottom are single-scale atomic vectors of different layers, while those on the right are multi-scale atomic vectors.}
\end{figure}

For predictive tasks shown in Fig. \ref{SL}, we randomly mask a certain proportion of nodes in a crystal graph during training to ensure that the elemental information of atoms corresponding to masked nodes is not available. The masked information is the period and the group of an element, following the conventions defined in the previous study\cite{XieT}, corresponding to classification problems of 9 categories for period and 18 categories for group. Besides, we can also randomly mask a certain proportion of edges connecting to masked atoms which encode the distance between atoms, to perform a classification problem of 41 categories. The selected 41-dimensional edge vectors are compeletely masked, meaning that all corresponding radial basis function (RBF) expansion coefficients of the chosen edges are concealed. The proportions of both node masking and edge masking are set to 0.15 in our work, but can be regarded as an additional hyperparamter to explore its effects to the self-supervised training in further studies. Then we train the GNN to reproduce the proper labels based on the edges and neighbouring nodes of the masked atoms. As a result, the model gradually learns the rules of chemical constituents of materials and can capture high-level information about local structures. The loss function of each classification problem in the predictive self-supervised learning is the cross-entropy loss function $L_{pred}$ implemented in Pytorch\cite{pytorch}, defined as
\begin{eqnarray}
L_{pred}& = &\log\left(\mathbf{z}, c\right) \nonumber 
= -z_{c} + \log\left(\sum_{j=1}^{M}\exp(z_{j})\right),
\end{eqnarray}
where the output vector $\mathbf{z}$ is an $M$-dimensional vector,  $M$ the number of categories in the classification problem, and $c$ the proper label. The definition of the total loss function in predictive self-supervised learning is the sum of the loss functions of two node-level and one edge-level classification tasks. 

For the GNN architecture, we first transform the node vector of length 92 and the edge vector of length 41 to the initial embeddings of length 64 by a linear layer as the preprocessing step. Then, we put the preprocessed crystal graph into a 5-layer CGCNN and update the node vectors by the graph structure through the message passing mechanism. We utilize the CGConv operator implemented by Pytorch-geometric\cite{FeyM} as convolutional operations,
\begin{equation}
\mathbf{x}_{i}^{'} = \mathbf{x}_{i} + \sum_{j\in N(i)}\sigma\left(\mathbf{z}_{i,j}\mathbf{W}_{f} + \mathbf{b}_{f}\right) \odot g\left(\mathbf{z}_{i,j}\mathbf{W}_{s} + \mathbf{b}_{s}\right)
\end{equation}
where $\mathbf{z}_{i,j}=[\mathbf{x}_{i}, \mathbf{x}_{j}, \mathbf{e}_{ij}]$ is the concatenation of the central atom $\mathbf{x}_{i}$, the neighboring atoms $\mathbf{x}_{j}$ and the edges $\mathbf{e}_{i,j}$ between them. $\mathbf{W}$ and $\mathbf{b}$ are learnable weights. $\sigma$ and $g$ are sigmoid and soft plus activation functions respectively. $\odot$ is element-wise product. Next, we perform a BatchNorm layer to stabilize the training process, speed up the convergence after each convolutional layer, and finally perform a ReLU nonlinear layer and a Dropout layer to enhance the expressive and generalization abilities.

\section{The performance of self-supervised atomic representation on magnetic moments}

We train a five-layer GNN model by self-supervised learning to capture the high-level information about elements and local structures of materials in the dataset. Once the GNN model is trained, we can generate 64-dimensional, single-scale node embeddings(NEs) for each material through a single forward pass shown in Fig. \ref{GE}. The single-scale descriptor generated from the ith GNN layer is  denoted as NEi, while NE0 refer to the initial embeddings. A multi-scale descriptor can be simply constructed by concatenating single-scale descriptors. NE01, for example, is a 128-dimensional vector from the concatenation of NE0 and NE1. It can be shown that multi-scale descriptors have overall lower cosine similarity, thus better distinguishability compared to single-scale descritors \cite{Supp}. We can further fine-tune the target properties with corrections from a larger spatial scale beyond NE01, and next generate an experimental magnetic moment dataset to obtain some physical insights and illustrate our strategy.

The magnetic moments in solids are different from the isolated magnetic moments due to the crystal field environment and the interactions between atoms. Therefore the study of the magnetic moments in solids provides an excellent platform to verify the effects of environmental information within the crystals at different ranges. Although magnetic moments, being a vectorial property, are hard to be approached by the CGCNN-based framework, the magnitude of magnetic moments is still a useful scalar quantity that determines the magnetic properties of materials, and it is a crucial step of high-throughput screening and designing of conventional magnetic materials\cite{PhamT,BalluffJ,HuebschM} and magnetic topological materials\cite{XuY,ChoudharyK}.

\begin{table*}[htbp]
	\caption{Prediction performance of the KRR model on the experimental magnetic moment dataset when OFM and different self-supervised atomic descriptors are taken as inputs, where the uncertainties are the standard deviations of test scores on five random test sets. The unit of the magnetic moment is $\mu_{B}$.}
	\centering
	\def\temptablewidth{0.92\textwidth}
	{\rule{\temptablewidth}{0.6pt}} 
	\begin{tabular*}{\temptablewidth}{@{\extracolsep{\fill}}ccccccccc}
		\hline
		&&OFM&NE0&NE1&NE01&NE012&NE01+OFM&\\
		\hline
		&MAE&0.860$\pm$0.107&0.997$\pm$0.086&0.805$\pm$0.066&\textbf{0.719}$\pm$0.054&0.726$\pm$0.043&0.800$\pm$0.070&\\
		\hline
		&MSE&1.477$\pm$0.230&1.798$\pm$0.281&1.536$\pm$0.224&\textbf{1.289}$\pm$0.226&1.318$\pm$0.235&1.371$\pm$0.241&\\
		\hline
		&R2&0.684$\pm$0.059&0.620$\pm$0.053&0.674$\pm$0.049&\textbf{0.727}$\pm$0.045&0.721$\pm$0.046&0.710$\pm$0.049&\\
		\hline
		&PCC&0.836$\pm$0.025&0.797$\pm$0.025&0.823$\pm$0.031&0.854$\pm$0.027&0.851$\pm$0.282&\textbf{0.855}$\pm$0.028&\\
	\end{tabular*}
	{\rule{\temptablewidth}{0.6pt}}
	\label{PP1}
\end{table*}

The dataset of experimental magnetic moments is generated from the experimental antiferromagnetic material database MAGNDATA\cite{GallegoS,SGallego}. Let us compare the best performing single-scale atomic descriptors and multi-scale atomic descriptors in predicting the local magnetic moments in solid materials. The prediction performance of NE0, NE3, NE4, NE5, NE0123, NE01234, NE012345 is also analyzed \cite{Supp}. It is worth noting that the performance of NE1 can already be comparable to the manually constructed atomic descriptor OFM, as shown in Table \ref{PP1}. Furthermore, the length of the vector representations of each GNN layer is only 64, much smaller than the 1056-dimensional vector representation of OFM, which means that we can achieve similar or slightly higher accuracy in predicting magnetic moments at a lower computational cost. The overall performance of multi-scale atomic descriptors is more satisfactory than that of a single GNN layer. We also find that NE01 concatenating with OFM can improve the performance of OFM alone. On the one hand, NE01 may contain more physical inductive bias, i.e., high-level information about elements and structures, increasing the expressive power of a single OFM; on the other hand, NE01, as a highly compact local atomic descriptor, only has a length of 128 in vector representation, and significantly improve the prediction ability of OFM. The MAE and MSE decrease 7.0\% and 7.2\% respectively, and the R2 and PCC increase 3.8\% and 2.3\% respectively, with a slight decrease of computational costs, which may contribute to the toolbox of feature engineering\cite{WardL}.  

\section{The analysis of self-supervised atomic representations and dimensional reduction} 

The model we trained on the computational material dataset can generate atomic representations for unknown experimental magnetic materials, whose overall sound performance shows that our self-supervised training strategy can indeed guide the model to learn the essential features of material data, i.e., rules of elements and structures.

To further characterize the rich information in the atomic representations generated by the self-supervised pre-trained GNN model, we utilize the non-linear, dimensional reduction method t-SNE\cite{VanM} to map the 128-dimensional NE01 atomic representations to two-dimensional (2D) space, visualizing and labeling these 2D points with element types.  The t-SNE plot of NE01 labeled with local environments can be found in Supplemental materials \cite{Supp}.

\begin{figure}[htbp]
	\centering
	\subfigure[]{
		\begin{minipage}[t]{0.4\linewidth}
			\centering
			\includegraphics[height=4.5cm,width=4.3cm]{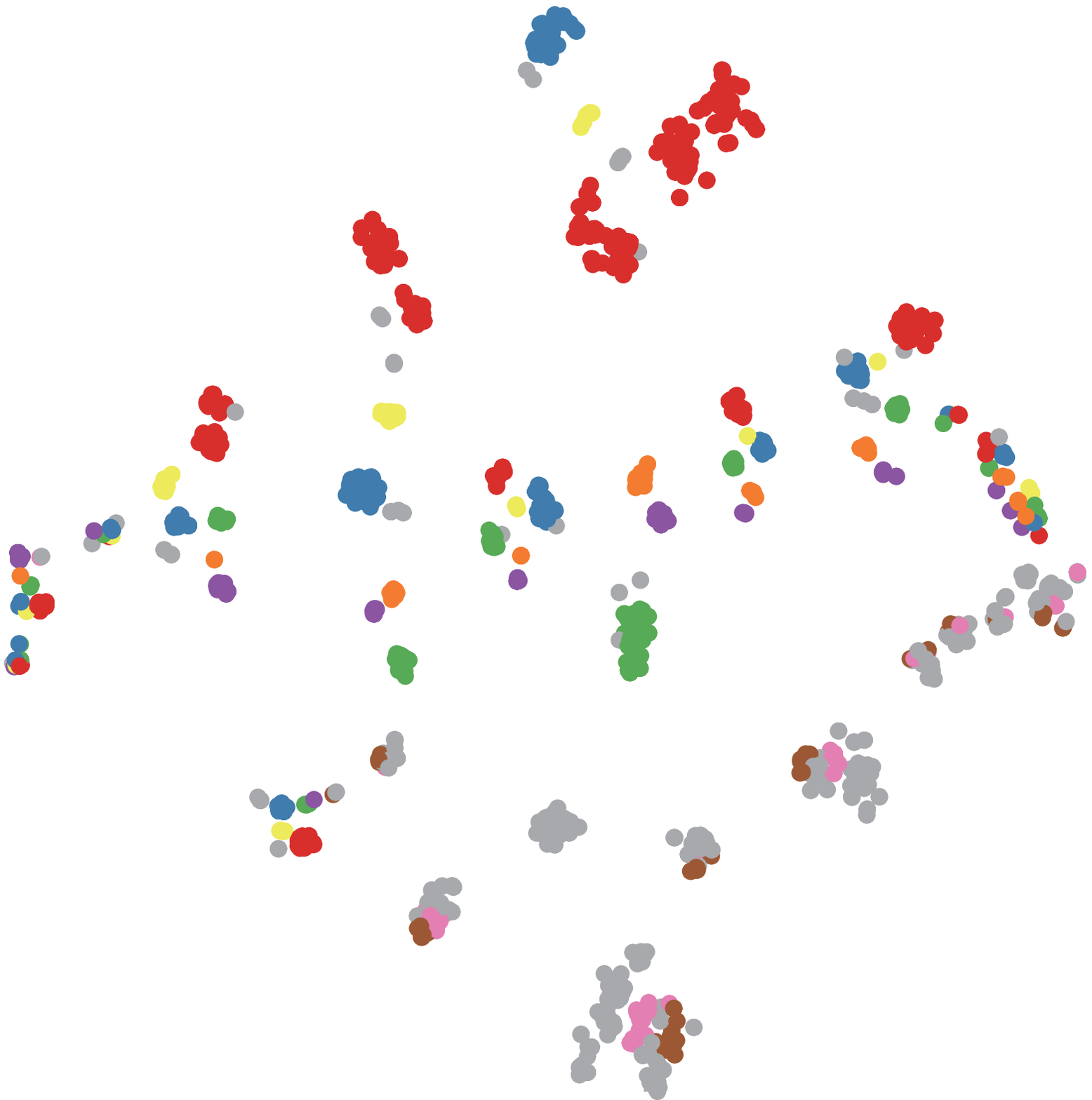}
			\label{RE}
	\end{minipage} }
    \hspace{.35in}
	\subfigure[]{
		\begin{minipage}[t]{0.4\linewidth}
			\centering
			\includegraphics[height=4.5cm,width=4.3cm]{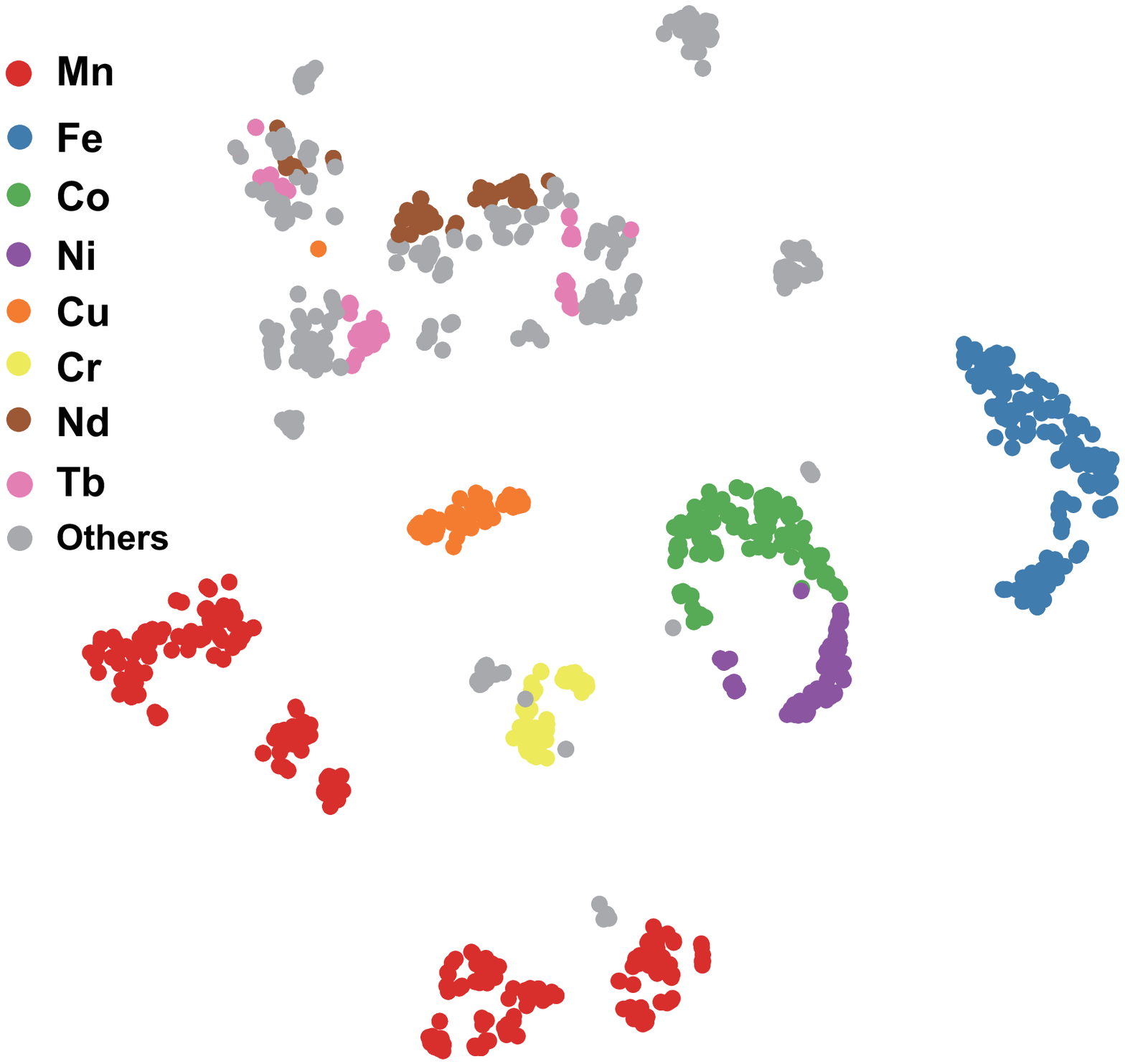}
			\label{SE}
	\end{minipage}}
	\caption{t-SNE visualization of self-supervised atomic vectors NE-SSL and random atomic vectors NE-Random with element labels, where the different colors represent different elements. (a) 2D distribution of NE-Random. (b) 2D distribution of NE-SSL.}
\end{figure}

The t-SNE visualizations of NE01 atomic representations NE-SSL and NE-Random are shown in Figs. \ref{RE} and \ref{SE}. The figures show that the atomic vectors NE-SSL generated by the self-supervised pre-trained GNN are better clustered under element labels, especially for transition metal elements shown in Fig. \ref{SE}. Lanthanides, such as Nd, Tb, and other grey-colored elements, are also clustered in the upper left region of Fig. \ref{SE}. In contrast, the random atomic vectors NE-Random show no clear organizational pattern in Fig. \ref{RE}, proving that self-supervised learning can capture rich chemical rules. From previous analysis, we can see the GNN can indeed learn the general rules of the elemental and structural distribution in materials. They keep the learned rules in the form of weights that are further transferred to the generated atomic representations. Therefore, NE-SSL has richer physical information than NE-Random, leading to performance improvements.

\section{The superiority of multi-scale atomic representation} 
\begin{figure}[htbp]
	\centering
	\subfigure[]{
		\begin{minipage}[t]{1.0\linewidth}
			\centering
			\includegraphics[height=4.0cm,width=8.0cm]{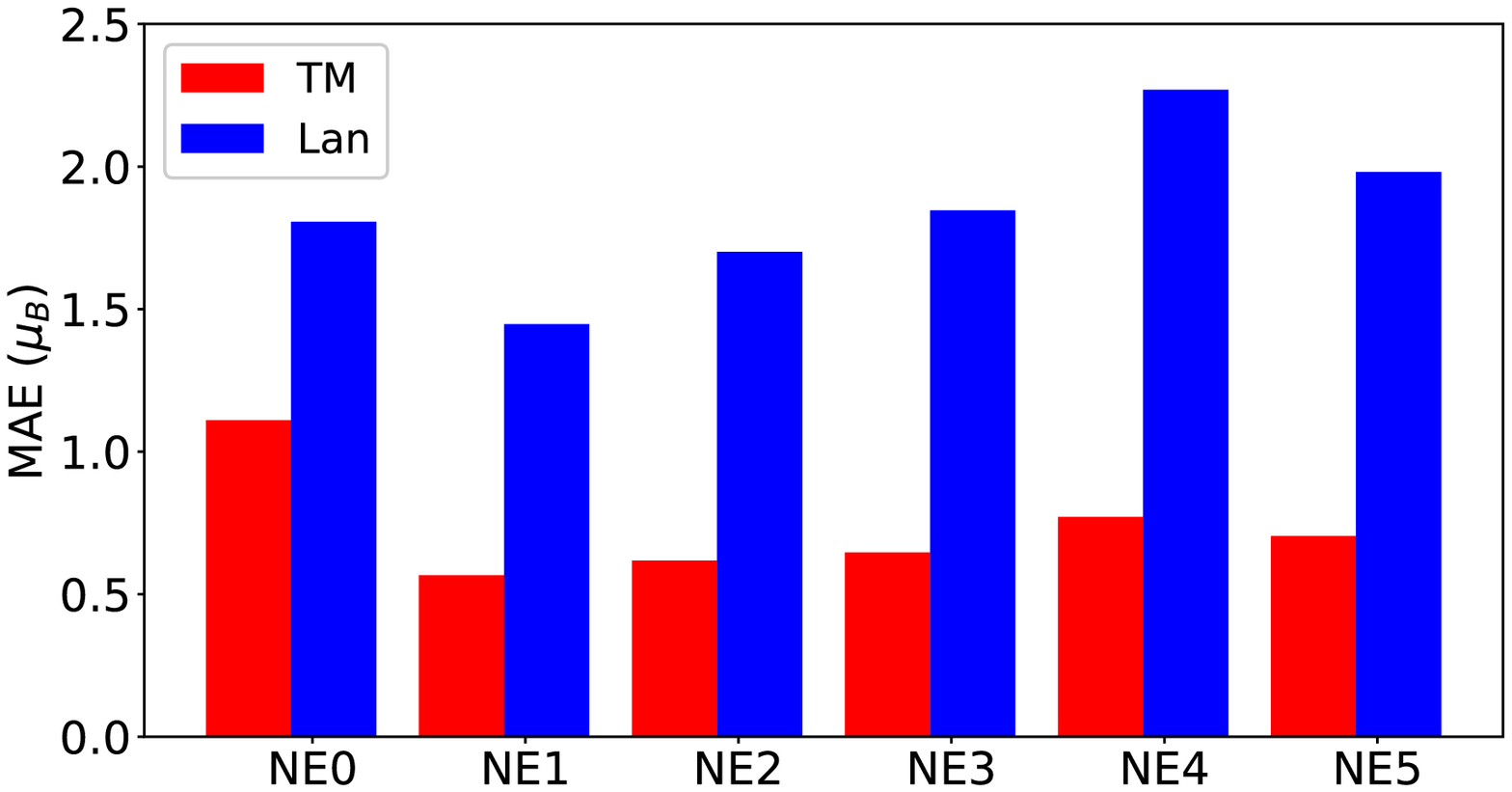}
			\label{TMLa1}
	\end{minipage} }
	\hspace{.20in}
	\subfigure[]{
		\begin{minipage}[t]{1.0\linewidth}
			\centering
			\includegraphics[height=4.0cm,width=8.0cm]{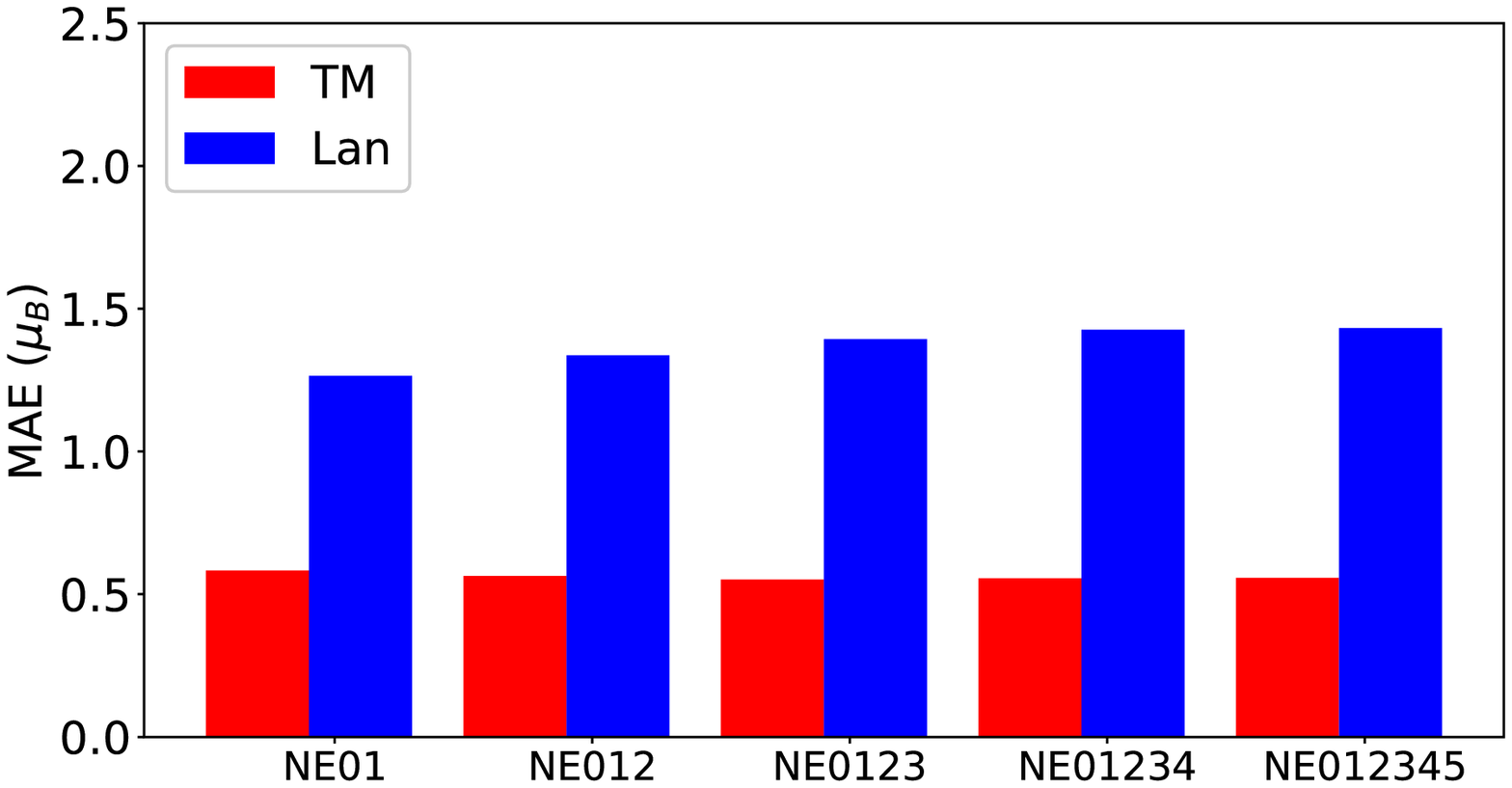}
			\label{TMLa2}
	\end{minipage}}
	\caption{The MAE of the KRR model on the magnetic moments of transition metal(red) and lanthanide(blue) in a test set given (a) single-scale and (b) multi-scale self-supervised atomic vectors as input.}
	\label{TMLa}
\end{figure}

To further analyze the results shown in Table \ref{PP1}, we pick out transition metal and lanthanide elements from a test set. The test set size is 255 for transition metal elements and 93 for lanthanide elements, respectively. We record the MAE of different atomic descriptors on the test set, shown in Fig. \ref{TMLa}. 

We can find that each atomic descriptor's prediction error of lanthanide elements is much larger than that of transition metal elements on the magnetic moment. The reason is twofold. (1) It may be caused by the different training data sizes of 1257 and 470 for transition metals and lanthanides, respectively. The small amount of lanthanides may lead to poor fitting in the training process. As a result, the machine learning model fails to learn the general rules about the magnetic moments of lanthanides compared to transition metals. (2) It may also result from the fact that the magnetic properties of lanthanides are more complex than transition metal elements. From NE0 to NE1, the prediction errors of transition metals and lanthanides drop significantly, which means the near-site environment already contains the relevant information about the local crystal field and spin-orbit coupling. For single-scale atomic representations from NE1 to NE4, the prediction errors of transition metals and lanthanides rapidly increase, indicating that atomic representations of the deeper GNN layer have a sustained loss of information at the local level, which corresponds to the over-smoothing problem; For multi-scale atomic descriptors from NE01 to NE012345, with the information of the larger spatial scales, the prediction errors of transition metals only slightly fluctuate and the prediction errors of lanthanides increase at a much smaller amount than those from NE1 to NE4. These results show that multi-scale atomic representations can preserve the essential information related to the magnetic moment and resist information loss and noises.

We can also see from Tables \ref{PP1} that the MSE is much greater than the MAE of each atomic descriptor, which indicates that there are outliers. So it is interesting to analyze the parity plots of predicted values versus the experimental values from the test set. We take the parity plot of NE01 as an instance to disclose more physical insights.

\begin{figure}[htbp]
	\centering
	\subfigure[]{
		\begin{minipage}[t]{1.0\linewidth}
			\centering
			\includegraphics[height=5.0cm,width=7.0cm]{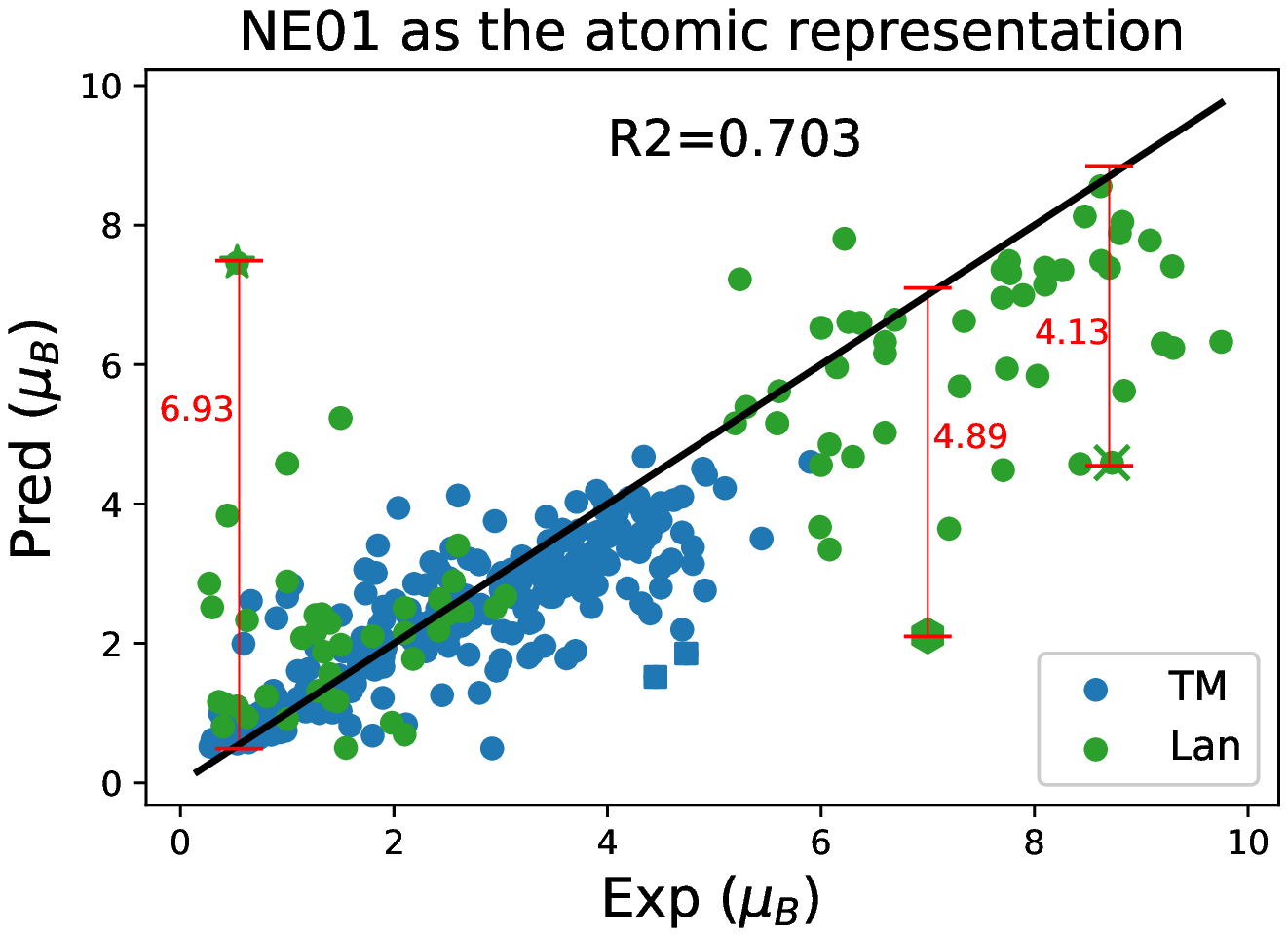}
		    \label{NE01}
		\end{minipage} 
	}
	\hspace{.20in}
	\subfigure[]{
		\begin{minipage}[t]{1.0\linewidth}
			\centering
			\includegraphics[height=5.0cm,width=7.0cm]{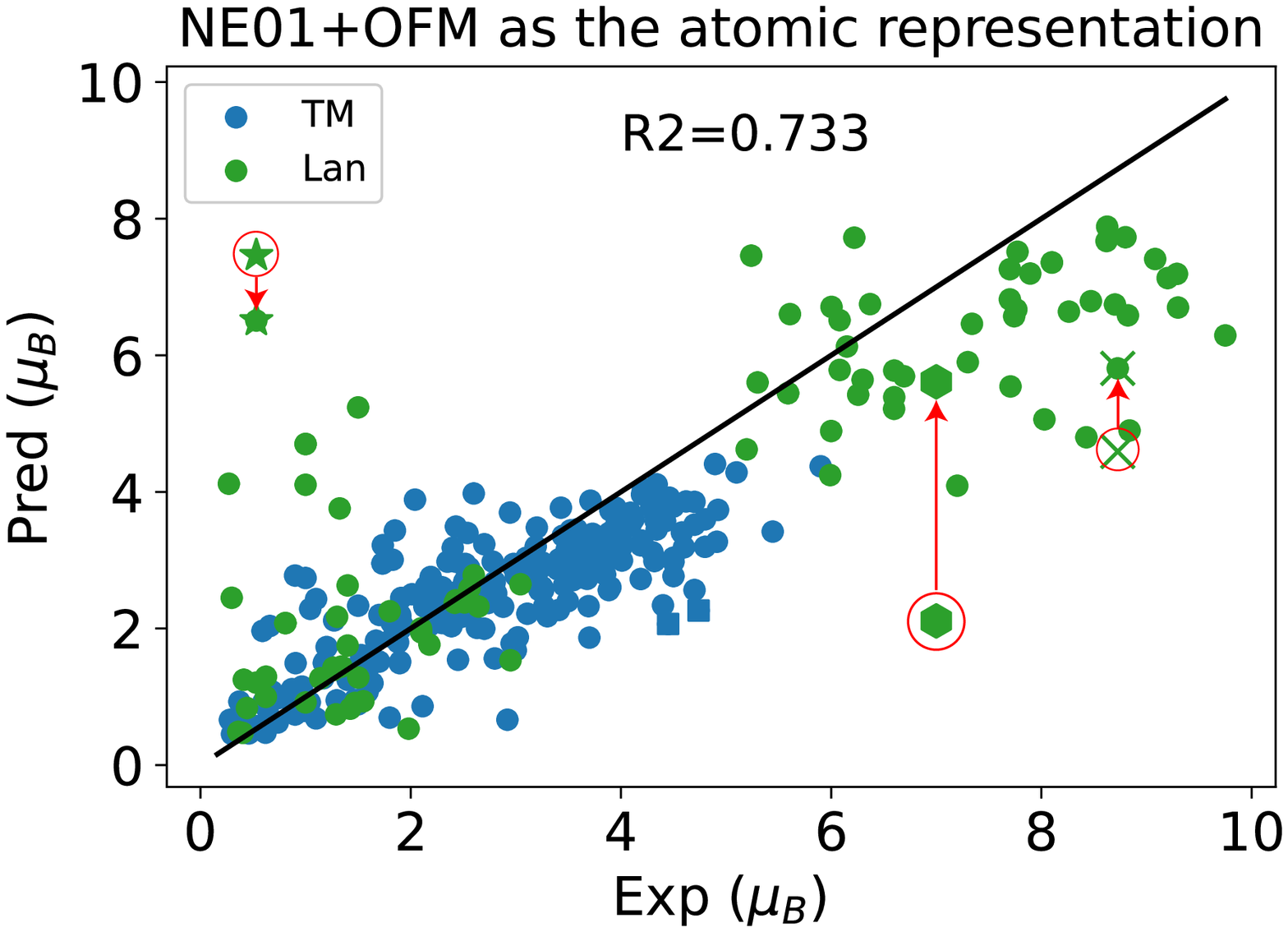}
			\label{NE01OFM}
		\end{minipage}
	}
	\caption{(a) Parity plot of the KRR model on a test set when the NE01 is taken as input, the green and blue points are transition metals and lanthanides respectively, the texts colored in red show prediction errors of the corresponding outliers. (b) Parity plot of the KRR model on a test set when the NE01+OFM atomic representation is taken as input, the red arrows indicate changes in prediction performance after incorporating OFM with NE01 compared to (a).}
	\label{NEtot}
\end{figure}

First, it is worth mention that in the test set of size 364, 52.2\% of elements satisfying $\vert \mathrm{Exp} - \mathrm{Pred}\vert \leqslant 0.5 \mu_{B}$, and 76.9\% of elements satisfying $\vert \mathrm{Exp} - \mathrm{Pred}\vert \leqslant 1.0 \mu_{B}$. In contrast, by the spin-density functional theory (SDFT), the errors of DFT values to experimental values within $0.5\mu_{B}$ and $1\mu_{B}$ are 51.9\% and 77.22\%, respectively\cite{HuebschM}. The SDFT calculation is also based on the MAGNDATA database, however, with much more computational resources. 

Next, we discuss the outliers in the parity plot shown in Fig. \ref{NE01}. (1) For transition metal elements, the magnetic moment with the most significant prediction error comes from the Mn atoms in Mn$_{6}$Ni$_{16}$Si$_{7}$\cite{AhmedSJ}, which are 2.88$\mu_{B}$ and 2.93$\mu_{B}$ respectively, shown as two blue squares located within $4 \mu_{B}<\mathrm{Exp}<5 \mu_{B}$ in Fig. \ref{NE01}. Mn$_{6}$Ni$_{16}$Si$_{7}$ is a material with rich physics in magnetism. For example, there is geometric frustration of antiferromagnetism in the octahedral structure formed by six Mn atoms below 197K. (2) For lanthanides, the most significant prediction error,  6.93 $\mu_{B}$, comes from the Dy atoms in Dy$_{3}$Ru$_{4}$Al$_{12}$\cite{GorbunovD}, shown as the green pentagon located within $0<\mathrm{Exp}<1 \mu_{B}$ in the upper left corner of Fig. \ref{NE01}. The non-collinear magnetic structure of the Dy$_{3}$Ru$_{4}$Al$_{12}$ is caused by the competition between the RKKY interaction and the anisotropy induced by the crystal field. (3) The prediction error of Tb atom in Tb$_{5}$Ge$_{4}$\cite{RitterC} is 4.13 $\mu_{B}$, shown as the green cross located within $8 \mu_{B}<\mathrm{Exp}<9 \mu_{B}$ in Fig. \ref{NE01}, which exhibits strange behaviors caused by spin reorientation transition of the canted antiferromagnetic structure. (4) The prediction error of Gd atoms in GdVO$_{4}$\cite{PalaciosE} is 4.89 $\mu_{B}$, shown as a green hexagon located at nearly $\mathrm{Exp} = 7 \mu_{B}$, which has high specific heat and strong magnetocaloric effect above 2.5 K. The four outliers mentioned above all have non-trivial magnetic structures or magnetic properties, indicating that the revealed materials with significant prediction errors have some curious physics and are worth further analysis.

Finally, we compare the parity plots of NE01 and NE01 concatenated with OFM (NE01+OFM), shown in Fig. \ref{NE01OFM}. We can see that the R2 score of NE01+OFM increases 4.3\% over NE01 alone with more explicit physical information involved in OFM, which means that the overall prediction performance improves. In particular, three lanthanides with the most significant prediction errors marked in Fig. \ref{NE01} are moving towards the experimental values. The significant improvement in predicting the magnetic moment of the Gd atom is similar to the SDFT calculation by adding Hubbard U in some lanthanides\cite{HuebschM}.

\section{The self-supervised material representation and the NEGNN framework} 
After analyzing the results of self-supervised atomic representations on the magnetic moment dataset, we turn to study various kinds of material properties. First, we treat the representations of different layers as environmental corrections at different spatial ranges and fine-tune the target properties by adjusting the environmental information included in the descriptors. On the one hand, like the manually constructed local atomic descriptors, we construct the graph embeddings (GE) of materials by simply averaging the atomic representations of the same material. They can be directly used as inputs of the KRR model to fit structure-property relations. On the other hand, the expressive ability of GNN can be enhanced by incorporating rich physical information or structural information\cite{KaramadM,BanjadeH} into the initial node vectors of original crystal graphs. Therefore, we introduce the NEGNN framework, which combines self-supervised atomic representations with graph structure of materials to perform end-to-end learning of specific properties. 

\begin{figure}[htbp]
	\centering
	\subfigure[]{
		\begin{minipage}[t]{1.0\linewidth}
			\centering
			\includegraphics[height=4.0cm,width=6.0cm]{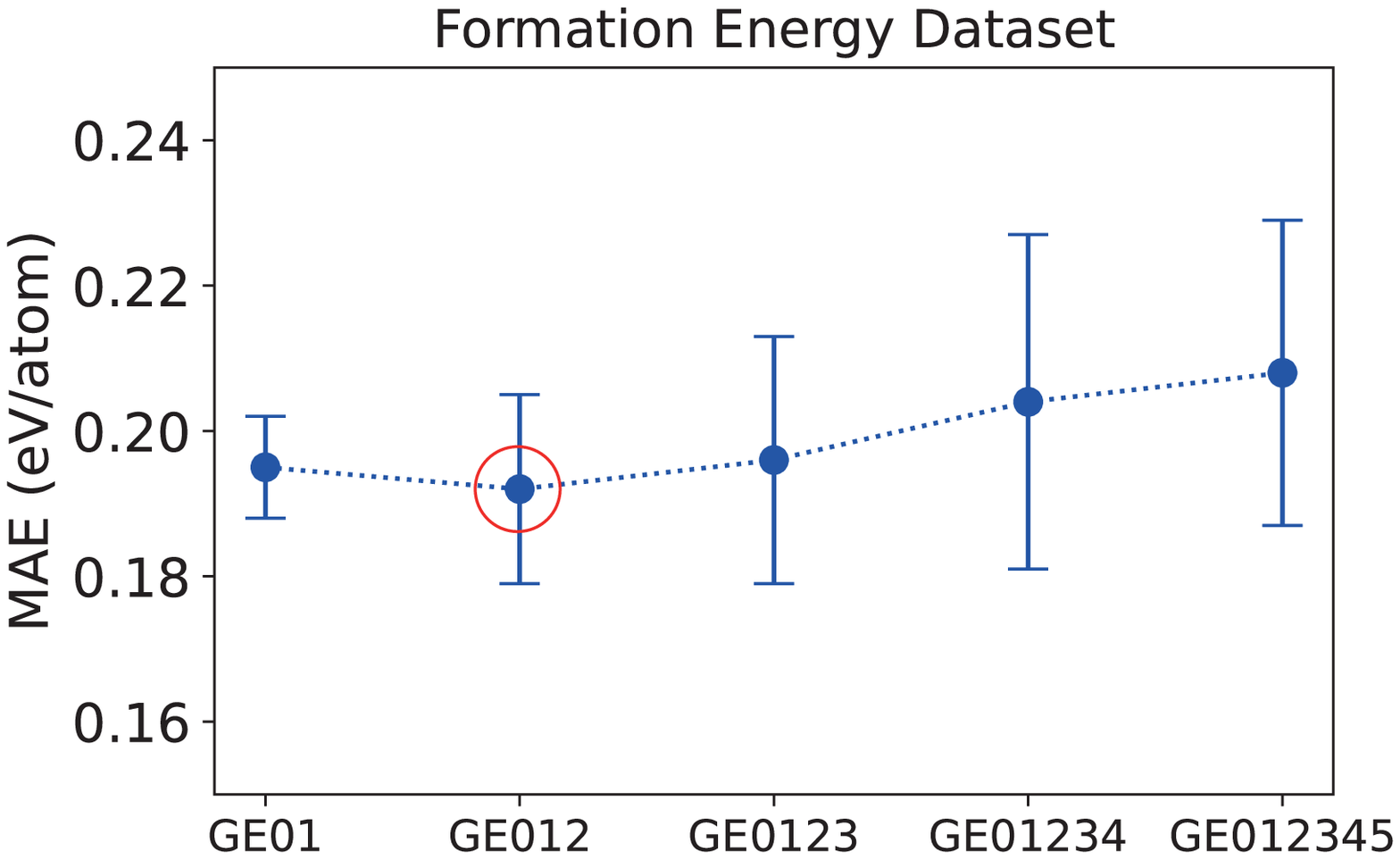}
			\label{Formation}
		\end{minipage} 
	}
	\hspace{.20in}
	\subfigure[]{
		\begin{minipage}[t]{1.0\linewidth}
			\centering
			\includegraphics[height=4.0cm,width=6.0cm]{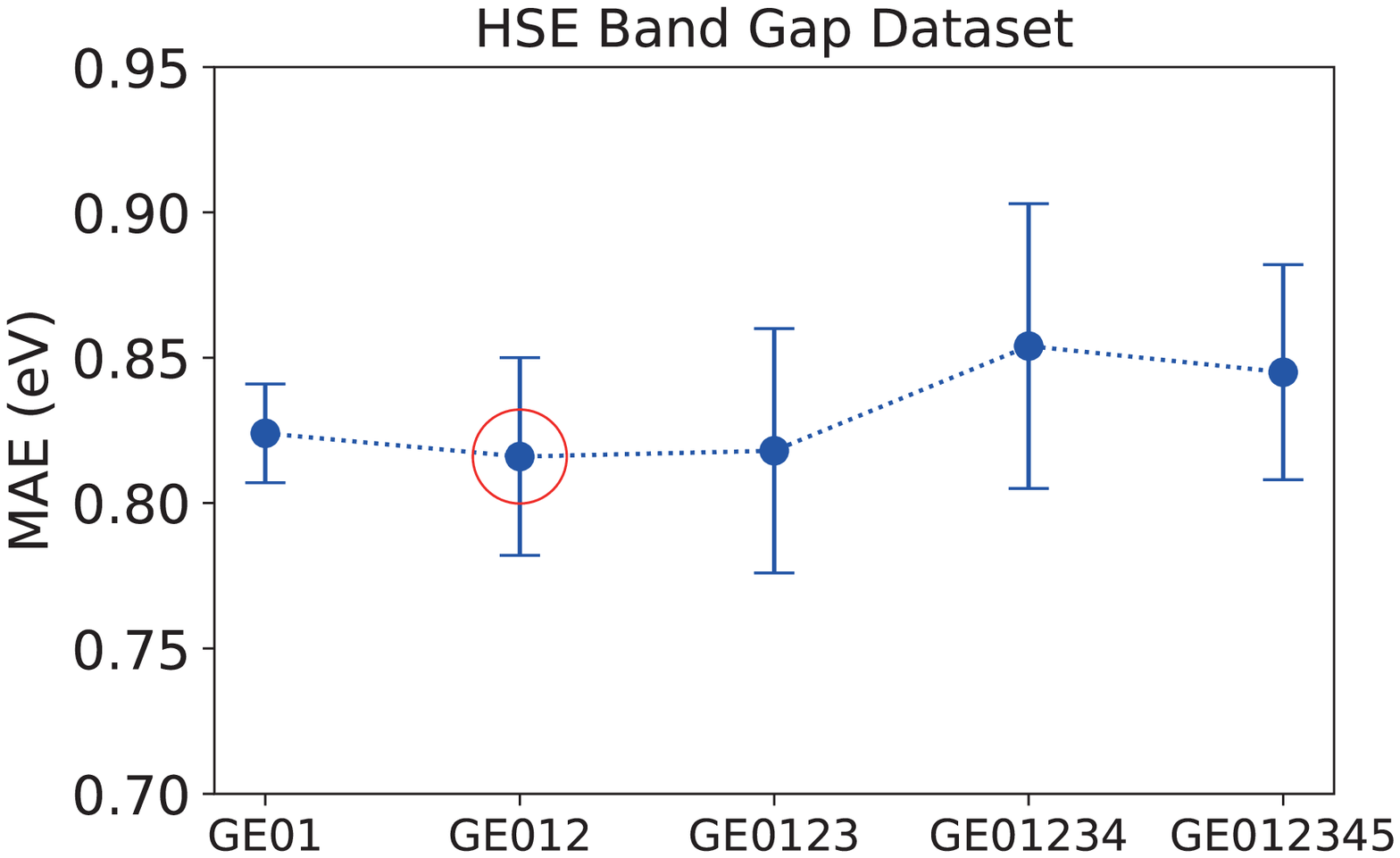}
			\label{Band}
		\end{minipage}
	}
	\caption{The MAE of the KRR model on the (a) formation energy dataset and (b) HSE bandgap dataset, when different multi-scale self-supervised material representations are taken as inputs. GE012 achieves the lowest prediction error indicated by the red circle.}
	\label{FormationBand}
\end{figure}

\begin{table*}[htbp]
	\caption{Prediction performance of the KRR model with different material descriptors on the formation energy dataset and HSE bandgap dataset. We choose the self-supervised multi-scale descriptors, GE012, with the best performance among the descriptors shown in Fig. \ref{FormationBand}. The units of formation energy and HSE band gap are eV/atom and eV respectively.}
	\centering
	\def\temptablewidth{0.8\textwidth}
	{\rule{\temptablewidth}{0.6pt}}  
	\begin{tabular*}{\temptablewidth}{@{\extracolsep{\fill}}cccccccccc}
		&&\multicolumn{5}{c}{Formation Energy}&\multicolumn{3}{|c}{HSE band gap}\\
		\hline
		&&ESM&ECM&SM&OFM&GE012&SM&OFM&GE012\\
		\hline
		&MAE&0.49&0.64&0.37&0.282$\pm0.018$&\textbf{0.192}$\pm$0.013&1.387$\pm$0.015&0.935$\pm$0.028&\textbf{0.816}$\pm$0.034\\
		\hline
		&MSE&---&---&---&0.158$\pm$0.017&\textbf{0.081}$\pm$0.010&3.464$\pm0.178$&1.587$\pm$0.101&\textbf{1.302}$\pm$0.061\\
		\hline
		&R2&---&---&---&0.857$\pm$0.014&\textbf{0.927}$\pm$0.005&0.305$\pm$0.040&0.680$\pm$0.036&\textbf{0.738}$\pm$0.020\\
		\hline
		&PCC&---&---&---&0.932$\pm$0.007&\textbf{0.966}$\pm$0.003&0.575$\pm$0.024&0.838$\pm$0.018&\textbf{0.863}$\pm$0.015\\
	\end{tabular*}
	{\rule{\temptablewidth}{0.6pt}}
	\label{ForHSE}
\end{table*}

To highlight the general information about elements and structures within the embeddings, we compare the performance of GE with several manually constructed descriptors and the CGCNN model on various material properties such as formation energy, HSE bandgap, and elastic properties. 

\begin{figure}[htbp]
	\centering
	\subfigure[]{
		\begin{minipage}[t]{1.0\linewidth}
			\centering
			\includegraphics[height=4.0cm,width=6.0cm]{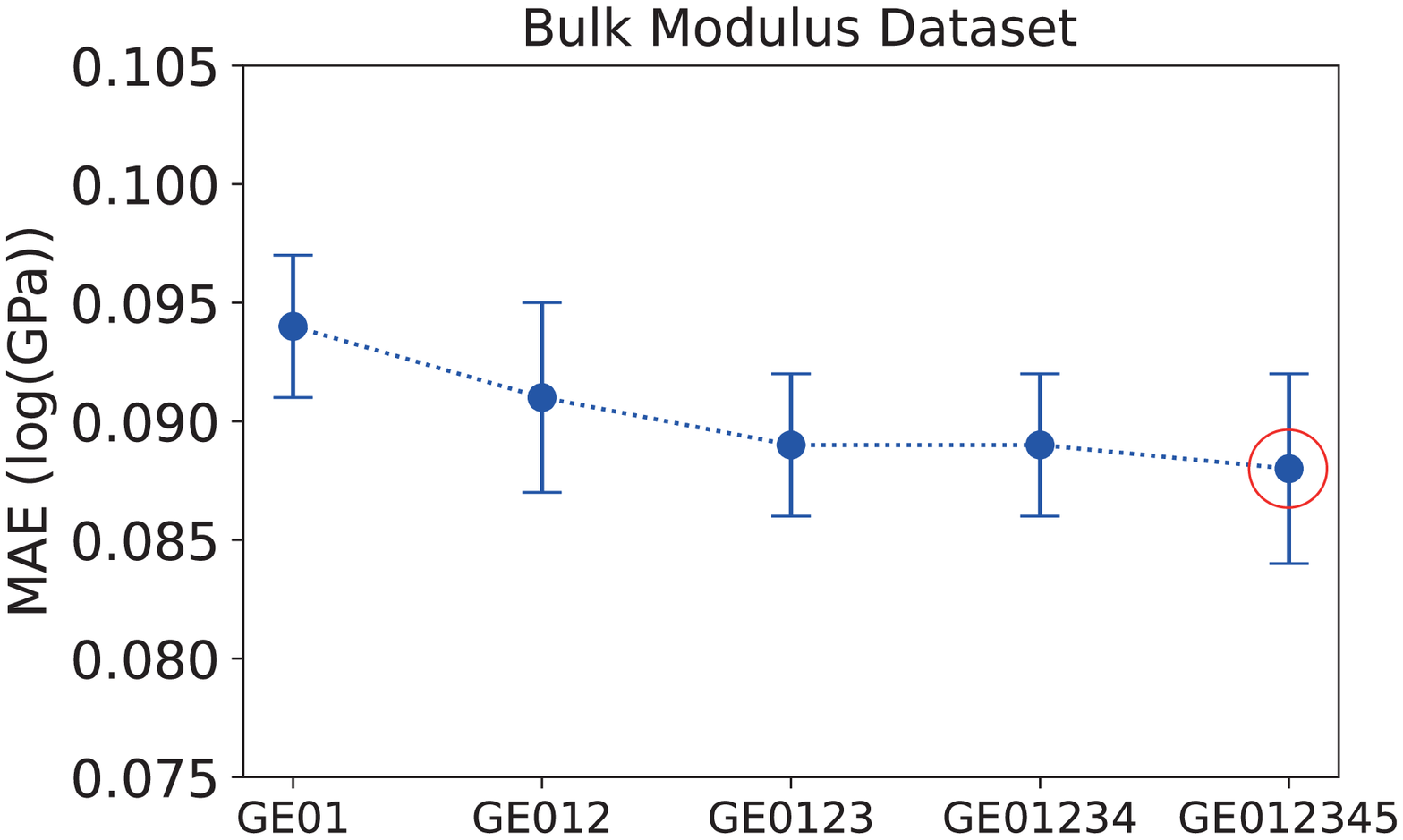}
			\label{Elastic1}
	\end{minipage} }
	\hspace{.30in}
	\subfigure[]{
		\begin{minipage}[t]{1.0\linewidth}
			\centering
			\includegraphics[height=4.0cm,width=6.0cm]{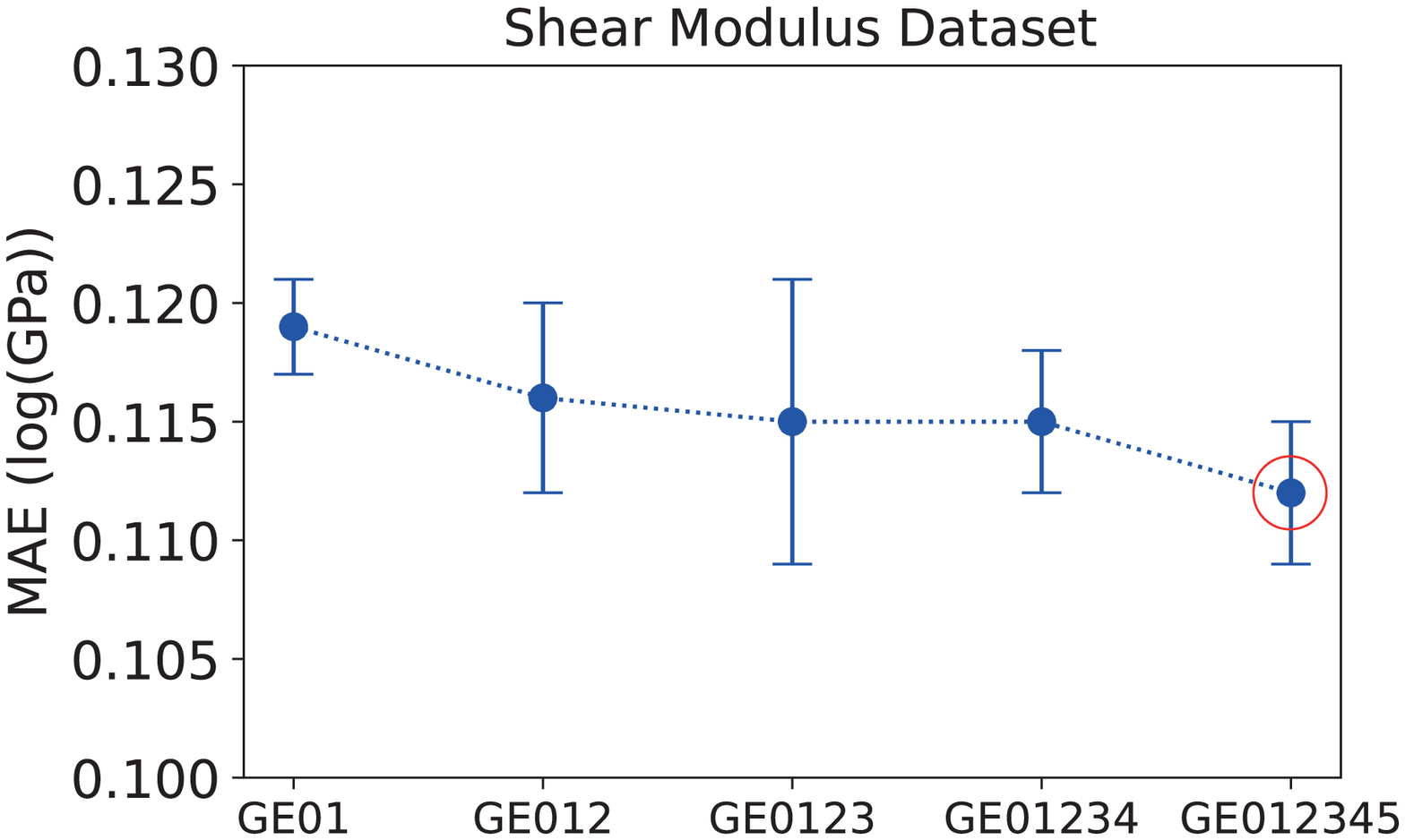}
			\label{Elastic2}
	\end{minipage}}
	\hspace{.30in}
	\subfigure[]{
		\begin{minipage}[t]{1.0\linewidth}
			\centering
			\includegraphics[height=4.0cm,width=6.0cm]{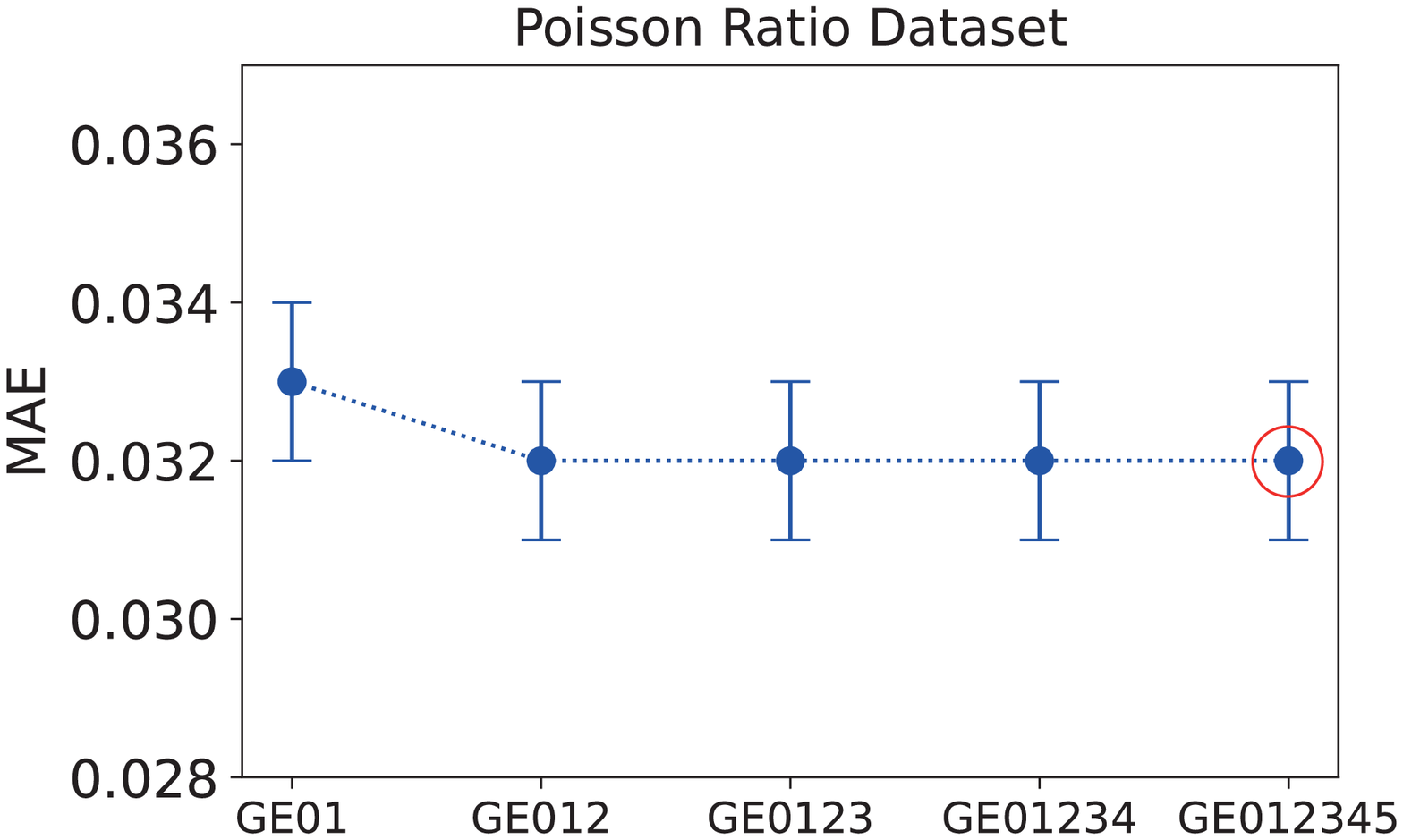}
			\label{Elastic3}
	\end{minipage}}
	\caption{The MAE of the KRR model on the (a) bulk modulus dataset, (b) shear modulus dataset, (c) Poisson ratio dataset when different self-supervised multi-scale material representations are taken as inputs. The units of bulk and shear modulus are taken as $\log(GPa)$. GE012345 achieves the lowest prediction error on all of the three elastic properties indicated by the red circle.}
	\label{Elastic}
\end{figure}

\begin{table*}[htbp]
	\caption{Prediction performance of the KRR model with different material descriptors on the three elastic property datasets. We choose the self-supervised multi-scale descriptors, GE012345, with the best performance among the descriptors shown in Fig. \ref{Elastic}. The units of bulk and shear modulus are taken as $\log(GPa)$.}
	\centering
	\def\temptablewidth{0.9\textwidth}
	{\rule{\temptablewidth}{1.0pt}}  
	\begin{tabular*}{\temptablewidth}{@{\extracolsep{\fill}}cccccccc}
		&&\multicolumn{2}{|c}{Bulk Modulus}&\multicolumn{2}{|c}{Shear Modulus}&\multicolumn{2}{|c}{Poisson Ratio}\\
		\hline
		&&OFM&GE012345&OFM&GE012345&OFM&GE012345\\
		\hline
		&MAE&0.121$\pm$0.017&\textbf{0.088}$\pm$0.004&0.142$\pm$0.011&\textbf{0.112}$\pm$0.003&0.034$\pm$0.001&\textbf{0.032}$\pm$0.001\\
		\hline
		&MSE&0.027$\pm$0.003&\textbf{0.017}$\pm$0.002&0.035$\pm$0.004&\textbf{0.024}$\pm$0.001&0.002$\pm$0.000&0.002$\pm$0.000\\
		\hline
		&R2&0.774$\pm$0.021&\textbf{0.860}$\pm$0.014&0.727$\pm$0.034&\textbf{0.817}$\pm$0.017&0.408$\pm$0.026&\textbf{0.471}$\pm$0.028\\
		\hline
		&PCC&0.894$\pm$0.006&\textbf{0.931}$\pm$0.008&0.868$\pm$0.007&\textbf{0.910}$\pm$0.012&0.648$\pm$0.016&\textbf{0.752}$\pm$0.015\\
	\end{tabular*}
	{\rule{\temptablewidth}{1pt}}
	\label{ElaT}
\end{table*}

First, we analyze the interplay between the range information within GE and the material properties. Fig. \ref{Formation} shows that the GE012 achieves the best performance among all multi-scale descriptors on formation energy, indicating that energy-related material properties can be well decomposed into the contributions from the local environment near the atom, consistent with the worse performance of the descriptors from GE0123 to GE012345 which has more extensive environmental information. It also holds for the bandgap dataset shown in Fig. \ref{Band} that the local environment plays the crucial role, which may result from the principle of electronic nearsightedness\cite{ProdanE}. While we can see from Figs. \ref{Elastic1} - \ref{Elastic3} that the descriptors incorporating the most extensive range of environmental information, namely GE012345, have achieved the best performance of all three elastic properties, indicating that the elastic properties are less local than formation energy and bandgap. Second, we compare the prediction performance of the self-supervised descriptors and the manually constructed descriptors shown in Tables \ref{ForHSE} and \ref{ElaT}. We can see that the self-supervised multi-scale descriptor, GE, is significantly better than the popular descriptors ESM, ECM, SM, and OFM, on all considered material properties. On the one hand, GE contain more general and high-level material information extracted through self-supervised learning instead of domain-specific information, such as Coulomb interaction encoded in SM and orbital interaction encoded in OFM. On the other hand, the compactness of GE reduces the possibility of overfitting and also results in lower computational costs.

\begin{table*}[htbp]\footnotesize
	\caption{Prediction performance of CGCNN and NEGNN on all the material properties.}
	\centering
	\def\temptablewidth{1.05\textwidth}
	{\rule{\temptablewidth}{1.0pt}}  
	\begin{tabular*}{\temptablewidth}{@{\extracolsep{\fill}}cccccccccccc}
		&&\multicolumn{2}{|c}{Formation Energy}&\multicolumn{2}{|c}{HSE Band Gap}&\multicolumn{2}{|c}{Bulk Modulus}&\multicolumn{2}{|c}{Shear Modulus}&\multicolumn{2}{|c}{Poisson Ratio}\\
		\hline
		&&CGCNN&NEGNN&CGCNN&NEGNN&CGCNN&NEGNN&CGCNN&NEGNN&CGCNN&NEGNN\\
		\hline
		&MAE&0.190$\pm$0.027&\textbf{0.180}$\pm$0.020&0.792$\pm$0.056&\textbf{0.705}$\pm$0.024&0.066$\pm$0.008&0.066$\pm$0.010&0.100$\pm$0.006&\textbf{0.096}$\pm$0.004&0.033$\pm$0.002&\textbf{0.029}$\pm$0.001\\
		\hline
		&MSE&0.081$\pm$0.010&\textbf{0.060}$\pm$0.011&1.452$\pm$0.658&\textbf{1.039}$\pm$0.104&0.012$\pm$0.003&\textbf{0.010}$\pm$0.002&0.017$\pm$0.001&0.017$\pm$0.001&0.002$\pm$0.000&0.002$\pm$0.000\\
		\hline
		&R2&0.933$\pm$0.019&\textbf{0.946}$\pm$0.008&0.729$\pm$0.085&\textbf{0.791}$\pm$0.028&0.898$\pm$0.021&\textbf{0.917}$\pm$0.015&0.863$\pm$0.008&\textbf{0.867}$\pm$0.012&0.467$\pm$0.032&\textbf{0.557}$\pm$0.027\\
		\hline
		&PCC&0.974$\pm$0.004&\textbf{0.980}$\pm$0.001&0.865$\pm$0.055&\textbf{0.897}$\pm$0.012&0.954$\pm$0.008&\textbf{0.965}$\pm$0.006&0.932$\pm$0.004&\textbf{0.940}$\pm$0.010&0.713$\pm$0.016&\textbf{0.752}$\pm$0.015\\
	\end{tabular*}
	{\rule{\temptablewidth}{1pt}}
	\label{GNNS}
\end{table*}

The NEGNN is distinct from the CGCNN, for that the input crystal graph is the combination of self-supervised multi-scale atomic representation and graph structure of materials as shown in Fig. \ref{pipeline}. We choose NE01 to combine with graph structure without loss of generality since it contains minimal sufficient information about elements and local environments. The prediction performance of CGCNN and NEGNN is shown in Table \ref{GNNS}.

Compared with the best performance of the KRR model in Tables \ref{ForHSE} and \ref{ElaT}, CGCNN shows improvement in accuracy yet loses computational efficiency on small datasets\cite{DunnA}. Nevertheless, the self-supervised enhanced GNN, NEGNN, improves the prediction performance on all material properties over CGCNN. The strength of self-supervised learning is from the ability to learn more high-level rules about materials. To check how the performance of self-supervised learning depends on the size of datasets, we evaluate both the NEGNN and CGCNN on the formation energy dataset (size 46,744) and bandgap dataset (size 27,430), denoted by "Formation Energy+" and "Bandgap+" in Table \ref{Larger}. These two datasets are taken directly from the original CGCNN paper\cite{XieT}, therefore we only train the NEGNN model and compare the prediction performance of NEGNN with that of the CGCNN from the published results. As we can see, NEGNN shows 13.78\% improvements over CGCNN in a larger bandgap dataset. In contrast, it achieves the same MAE as the CGCNN on an even larger formation energy dataset, indicating that the self-supervised learning strategy preserves its strength for the larger datasets.

\begin{table}[htbp]
	\caption{Prediction performance of CGCNN and NEGNN on formation energy dataset and bandgap dataset of larger size}
	\centering
	\def\temptablewidth{0.4\textwidth}
	{\rule{\temptablewidth}{1.0pt}}  
	\begin{tabular*}{\temptablewidth}{@{\extracolsep{\fill}}cccccc}
		&&\multicolumn{2}{|c|}{Formation Energy+}&\multicolumn{2}{c}{Bandgap+}\\
		\hline
		&&CGCNN&NEGNN&CGCNN&NEGNN\\
		\hline
		&MAE&\textbf{0.039}&\textbf{0.039}&0.388&\textbf{0.341}\\
		\hline
		&MSE&---&0.005&---&0.325\\
		\hline
		&R2&---&0.995&---&0.854\\
		\hline
		&PCC&---&0.998&---&0.924\\
	\end{tabular*}
	{\rule{\temptablewidth}{1pt}}
	\label{Larger}
\end{table}

\section{The comparison of the NEGNN and other popular GNN frameworks} 

NEGNN's main focus is to incorporate as much pertinent material data as possible at the beginning stage via two fresh methods. On the one hand, it can be a practical technique for recovering the randomly masked material properties to input essential information in a self-supervised manner; on the other hand, multiscale representations can also integrate information at different physical ranges and enhance the distinguishability between atomic vectors.

At this point,  it is necessary to highlight the differences between various GNN frameworks. The comparison between the original CGCNN\cite{XieT}, the primary reference of our work, and the NEGNN has already been discussed in detail. The SchNet\cite{SaucedaH} is one of the early GNN frameworks in material science, which utilizes a specific-designed continuous filter as the message-passing layer, primarily suitable for modeling small molecular systems. The MEGNet\cite{ChenC} proposed by Chen et al. is another early GNN framework. The MEGNet unifies the property prediction of crystals and molecules and incorporates a global state vector for specific purposes. However, the differences between MEGNet and NEGNN are distinctive: (a) The learned embeddings in MEGNet come from the supervised training process, which benefits from labeled data. In contrast, the NEGNN generates embeddings from self-supervised training, requiring only the primitive structure information of crystals. The NEGNN may benefit when the labels are computationally expensive or hard to acquire. (b) Regarding interpretability, both works provide the t-SNE plot of elemental embeddings and the rediscovery of chemical rules. Nevertheless, the NEGNN offers additional degrees of freedom to extract physical information, \textit{i.e.}, the scale of embeddings. The AtomSets\cite{chen2021atomsets} is a newly developed framework that aims to utilize the pre-training of MEGNet to generate universal feature vectors and improve prediction performance on small datasets. However, there are also several significant differences between AtomSets and NEGNN: (a) For the pre-training strategy, AtomSets performs transfer learning using large, labeled datasets, while our self-supervised approach does not need any labeled datasets; (b) The NEGNN utilizes multiscale embeddings with lower similarity and can gain in-depth physical information and better interpretability by adjusting the scale degree of freedom and analyzing the changes in the test dataset. For instance, we can verify the importance of local environments to both transition metal elements and lanthanide elements in Fig. \ref{TMLa}; we can identify some interesting magnetic materials with non-trivial magnetic behavior through the parity plot in Fig. \ref{NEtot};  we can also get a measure of the locality of target material properties in Fig. \ref{FormationBand} and Fig. \ref{Elastic}. All of the physical information is available by multiscale embeddings. Besides the two early works, we note that a Crystal Twins\cite{Magar2022} model was published after our submission. The Crystal Twins leverage a contrastive training strategy based on CGCNN with complicated loss functions. However, the cross-entropy loss functions in NEGNN are more transparent for physical information.

The superiority of NEGNN over CGCNN in accuracy can be seen from TABLE \ref{GNNS} and TABLE \ref{Larger}. The MEGNet shows MAEs of 0.032 eV/atom and 0.35 eV for formation energy and band gap\cite{ChenC}, using dataset sizes similar to those in TABLE \ref{Larger}. The performance of MEGNet can be competing with NEGNN.  However, only the NEGNN can survive the computationally expensive or unavailable labels. We have also tested the state-of-the-art framework, the ALIGNN\cite{Choudhary2021}, on the same datasets. It achieved 0.028 eV/atom and 0.275 eV, which is much better than CGCNN and NEGNN, and this may be attributed to incorporating more high-level material information. Notwithstanding, our findings revealed that ALIGNN's computational cost is substantially higher than NEGNN's since it requires additional GNN layers to update the atomistic line graph expressing the three-body interactions. This additional depth increases the number of training parameters, thus significantly increasing computation time. NEGNN still has the potential to improve by enlarging the unlabeled, pretraining dataset compared to all the frameworks mentioned above.

\section{Conclusion and outlook}

We introduce low-dimensional, fixed-length atomic representations by self-supervised learning on crystal graphs. Combining self-supervised descriptors with standard machine learning models such as KRR can predict atomic properties and several material properties more accurately than popular manually constructed descriptors like OFM and SM while maintaining good computational efficiency. A standard machine-learning model with self-supervised atomic representations is more trouble-free to train than the GNN model. It can avoid the overfitting problem usually suffered by the GNN models on small datasets. By altering the range of environmental information in the self-supervised atomic representation, we can gain a machine-learning model with good physical insights. The predictive self-supervised pre-training strategy can extract high-level information from unlabeled datasets and incorporate prior information of materials into the GNN model. The multi-scale embeddings can extract in-depth physical information with better interpretability. Based on the strategy of self-supervised learning and the generated multi-scale embeddings, we develop the knowledge-enhanced NEGNN framework by combining the self-supervised atomic vectors with the GNN, significantly improving the performance. 

The NEGNN framework is promising for various applications and is open for further developments. First, more self-supervised learning tasks can be performed on crystal graphs by encoding explicit structural information like bond angles, which can capture the more high-level information of materials and transfer it to atomic representations; Second, self-supervised learning can capture the local structures by recovering the distance between atoms, and the GNN can be further regularized by reproducing global structural information like crystal groups, which may be more potent in predicting material properties; Third, we have demonstrated that the effectiveness of self-supervised learning in predicting the atomic-level properties, especially for magnetic moments. It is not difficult to generalize the self-supervised atomic representation for other site-related properties in solid materials, for example, implementing more powerful machine learning models for impurity levels\cite{MannodiK} or bandgap engineering\cite{SharmaV,FreyN}.

$ $

\begin{acknowledgments}
This work has been supported by the research foundation of Institute for Advanced Sciences of CQUPT
(Grant No. E011A2022328), the Strategic Priority ResearchProgram of the Chinese Academy of Sciences (Grant No. XDB0460000) and the Innovation Program for Quantum Science and Technology (Grant No. 2024ZD0300104).
\end{acknowledgments}

\section*{DATA AVAILABILITY}

The code and all the datasets of this paper can be available from the GitHub \cite{link}.





\newpage

\end{document}


\captionsetup[figure]{name={Supplementary Figure}}
\captionsetup[table]{name={Supplementary Table}}

\begin{center}
\LARGE{Supplemental Material for Self-supervised Representations and Node Embedding Graph Neural Networks for Accurate and Multi-scale Analysis of Materials}
\end{center}


\newpage

\section{The generation of atomic representations and over-smoothing problem in deep GNN} 

 Through the message passing mechanism of GNN, the atomic representations of each layer are supposed to have some information about previous layers. However, it may not be ideal to take the atomic representation of the last layer as the final atomic representation in the trained GNN model. On the one hand, deeper GNNs can capture long-range interactions and extract more abstract and complex representations; on the other hand, deep GNNs often suffer the over-smoothing problem\cite{OonoK,LiQ}. More specifically, a $K$-layer GNN model means that nodes are updated by aggregating the information from their K-hop neighborhoods. As the number of layers increases, the receptive fields of different nodes will overlap significantly so that the final representations of nodes tend to converge to the same vector. For a CGCNN model, the over-smoothing problem may be more severe for materials with less atoms in the primitive cell.

\begin{figure}[htbp]
	\centering
	\includegraphics[height=5.0cm,width=8.0cm]{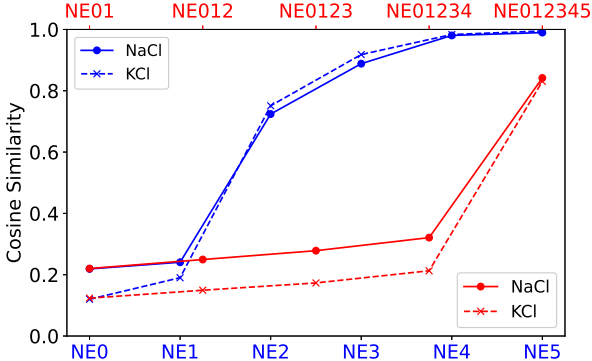}
	\caption{Cosine similarity between atomic vectors of Na (K) atoms and Cl atoms in NaCl (KCl) primitive cell under different GNN layers. The blue(red) lines correspond to single(multi)-scale atomic vectors.}
	\label{similarity}
\end{figure}

An excellent atomic descriptor should be able to distinguish between atoms located in different crystal environments\cite{GhiringhelliL,PozdnyakovS}. We define a simple indicator, cosine similarity, as the similarity measure of atomic representations. The cosine similarity of the vector $\mathbf{x}_{1}$  and $\mathbf{x}_{2}$ is defined as

\begin{eqnarray}
cosine\;similarity = \dfrac{\mathbf{x}_{1}\cdot\mathbf{x}_{2}}{x_{1}\times x_{2}} 
\end{eqnarray}

We take the two-atom primitive cells of NaCl and KCl as examples, for which are expected with over-smoothing problems. After putting the corresponding crystal graphs into the five-layer self-supervised pre-trained GNN, we can get the node embeddings (NE) of Na (K) atoms and Cl atoms of five different GNN layers. The cosine similarity between Na (K) and Cl embeddings under the same GNN layer is calculated to indicate the degree of over-smoothing, the results are shown in Supplementary Fig. \ref{similarity}. We can see that the similarity increases rapidly for single-scale descriptors(blue line), starting from the second GNN layer. The atomic representations of the last layer reach the highest similarity, \textit{i.e.} the representation of Na (K) atom is almost indistinguishable from Cl atom, and the atomic representations of the initial and first layers bear the lowest similarity, which indicates that the atomic representations of the last GNN layer significantly lose their local, atomic-level information. The multi-scale atomic representations(red line) have overall better distinguishability, as shown in Supplementary Fig. \ref{similarity}.

\newpage

\section{The datastet and the training details} 

\begin{figure}[htbp]
	\centering
	\subfigure[]{
		\begin{minipage}[t]{0.4\linewidth}
			\centering
			\includegraphics[height=4.8cm,width=6.0cm]{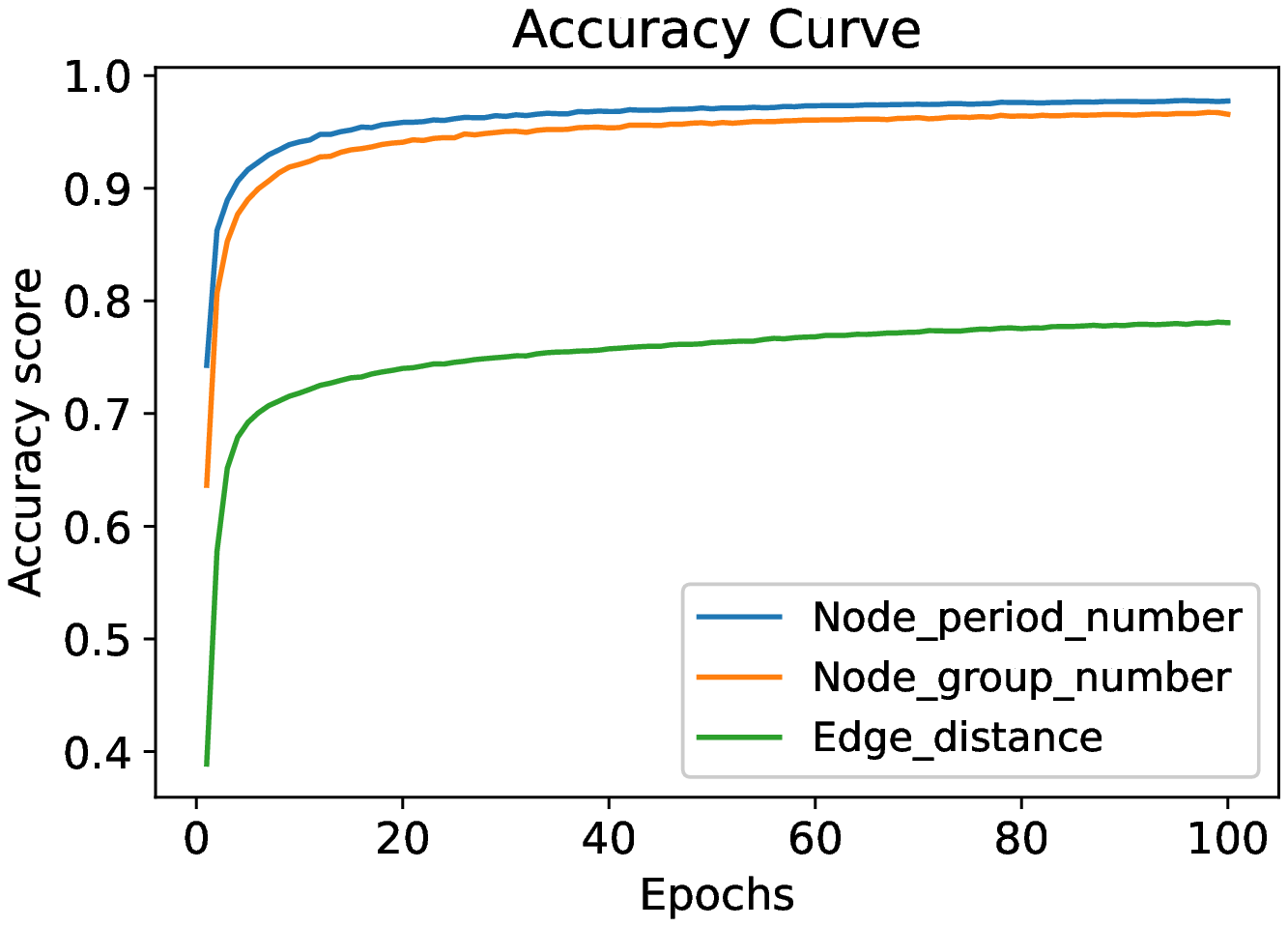}
			\label{acc_avg}
	\end{minipage} }
	\hspace{.40in}
	\subfigure[]{
		\begin{minipage}[t]{0.4\linewidth}
			\centering
			\includegraphics[height=4.8cm,width=6.0cm]{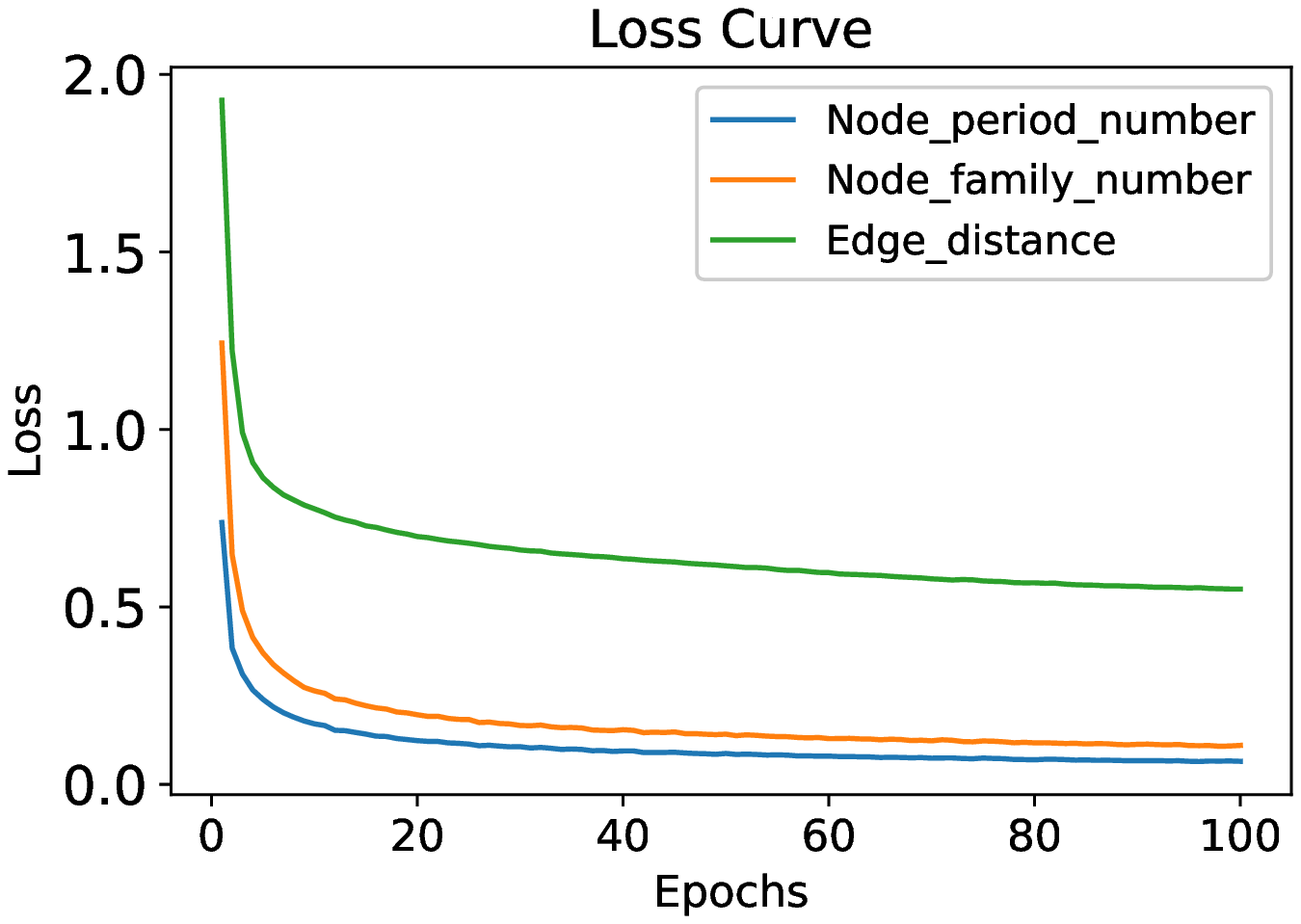}
			\label{loss_avg}
	\end{minipage}}
	\caption{(a) The training accuray vs epoch plot for predictive self-supervised learning. (b) The training loss vs epoch plot for predictive self-supervised learning.}
	\label{acc_loss}
\end{figure}

The learning curves of the self-supervised learning task are shown in Supplementary Figs. \ref{acc_loss}. We can see that the learning curves of the predictive task in Supplementary Fig. \ref{acc_avg} and Supplementary Fig. \ref{loss_avg} converge to different values depending on the number of categories in the classification task. The number of categories corresponding to the node is relatively small, 9 and 18 for period and group numbers, respectively, therefore the prediction accuracy can achieve almost 1. On the other hand, the number of categories corresponding to the edge is relatively large, 41 categories, and the accuracy eventually reaches nearly 80\%. All the training processes converge after 100 epochs.

The magnetic moment dataset is generated from the experimental magnetic material database MAGNDATA. The MAGNDATA database contains more than 1,500 experimental magnetic structures; after removing the magnetic structures that cannot be processed by Pymatgen\cite{ShyueP}, such as disordered magnetic structures and incommensurate magnetic structures, there remain 1,137 magnetic materials, covering 79 elements in total. We can generate a total number of 13,276 atoms with nonzero magnetic moments from these 1,137 magnetic materials, but we only retain atoms with different magnetic moments in the same material to avoid possible duplicate data. The remaining magnetic moment dataset of size 1816 covers 29 elements, including 13 transition metal elements (Co, Rh, Ni, V, Mn, Mo, Fe, Re, Os, Cu, Ir, Cr, Ru), 12 lanthanides (Ho, Ce, Tm, Dy, Pr, Tb, Nd, Yb, Eu, Gd, Er, Sm), and four other elements (Ga, Np, U, Pu), where Fe, Mn, Co, and Ni has the largest number of magnetic moments among all the elements.

The formation energy, HSE bandgap, and elastic properties dataset: (1) We choose the formation energy dataset from Ref. \cite{FaberF} and the corresponding 3,725 material structures from the Materials Project. The MAE reported in the Ref. \cite{FaberF} for Ewald sum matrix (ESM), extended Coulomb matrix (ECM) and SM are 0.49eV/atom, 0.64eV/atom and 0.37eV/atom respectively. (2) We choose the HSE bandgap dataset from the MaterialGo website\cite{JieJ} and the corresponding 6,030 material structures from the Materials Project. To perform the regression task, we only hold the 2,711 materials labeled by non-zero bandgap and with no more than 50 atoms in a unit cell for a reasonable length of the matrix-based descriptors such as SM, CM, and ESM. (3) We choose the elastic property dataset from the Ref. \cite{XieT} and the corresponding 3,204 material structures from the Materials Project, which are labeled by three kinds of elastic properties: bulk modulus, shear modulus, and Poisson ratio. 

\begin{table*}[htbp]
	\caption{The hyperparameters of KRR and CGCNN.}
	\centering
	\def\temptablewidth{1.052\textwidth}
	{\rule{\temptablewidth}{0.5pt}}  
	\begin{tabular*}{\temptablewidth}{@{\extracolsep{\fill}}ccccc}
		\hline
		&Hyperparameters of KRR&Range&Hyperparameters of CGCNN&Range\\
		\hline
		&kernels&['laplacian', 'rbf']&Learning rate&[1e-5, 1e-4, $\cdots$, 1e-1]\\
		\hline
		&kernel width $\gamma$&[1e-10, 1e-9, $\cdots$, 1e-0]&The number of GNN layers&[2,3,4,5]\\
		\hline
		&regularization strength $\alpha$&[1e-10, 1e-9, $\cdots$, 1e-0]&The number of hidden layers&[1,2,3,4]\\
		\hline
		&&&Dimension of node vectors in GNN&[32, 64, 96, 128] \\
	\end{tabular*}
	{\rule{\temptablewidth}{0.6pt}}
	\label{hyperpara}
\end{table*}

We utilize conventional machine learning models such as kernel ridge regression(KRR) for magnetic moment dataset, which is believed to be more efficient and accurate than neural networks on smaller dataset\cite{DunnA}. We split the dataset into a training set and a test set at a ratio of 8:2. Then we perform 10-fold cross-validation and grid search on the training set to determine the best hyperparameters of the regression model, and apply them to the test set. The hyperparameters of KRR as shown in Supplementary Table \ref{hyperpara} are kernels such as RBF (radial basis function) kernel or laplacian kernel, where the kernel width is $\gamma$, and the regularization strength is $\alpha$. This cross-validation procedure usually gives a more reasonable evaluation for small datasets\cite{DunnA}. We retrain the model based on the optimal set of hyperparameters and obtain the scores on the test set. We repeat the previous procedure five times using different train-test splits to get a reasonable estimate of model performance on a small dataset. The final score is averaged over five test scores, and the standard deviation over five test scores is taken as uncertainty. However, for a dataset more prominent than 10,000, we get a final score on the test set by a single train-validation-test split. The metrics for evaluating the prediction performance of the model are mean absolute error (MAE), mean squared error(MSE), R2 score, and Pearson correlation coefficient (PCC). The inputs are the orbital field matrix (OFM) and the atomic representations of different GNN layers and their concatenations, i.e., NE0, NE1, ..., NE01, ... All the inputs are fixed-length vectors.

For CGCNN and NEGNN, the hyperparameters shown in Table \ref{hyperpara} are the learning rate, the number of GNN layers, the dimension of node vectors within a GNN layer, and the number of hidden layers. The dimension of the hidden layer is taken as twice that of node vectors in GNN, which is usually a good choice. We also choose Adam as the optimizer for simplicity. 

\newpage

\section{The prediction performance on magnetic moments}

We have tested the performance of self-supervised atomic vectors on magnetic moments with several other popular ML models, such as the Support Vector Regression (SVR) and Random Forest (RF) regression in addition to the KRR model, as shown in Supplementary Fig. \ref{CompML}. We found that the SVR and KRR models achieve similar prediction performances, while the RF model performs the worst among them. It should also be noted that the prediction performances of the three ML models share a similar trend from NE0 to NE5, which indicates the robustness of self-supervised vectors against their scales.

\begin{figure}[htbp]
	\centering
	\includegraphics[height=4.8cm,width=7.0cm]{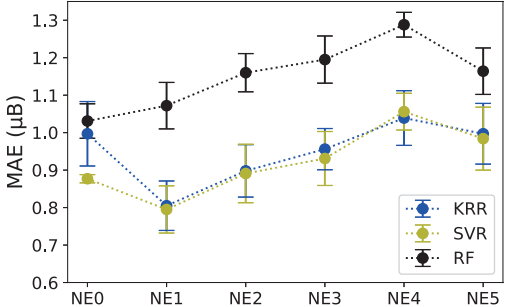}
	\caption{The prediction MAE of the KRR model, the SVR model and the RF model on the magnetic moments when NE0, NE1, ..., NE5 are taken as inputs vectors.}
	\label{CompML}
\end{figure}

\begin{table*}[htbp]
	\caption{Prediction performance of the KRR model on the experimental magnetic moment dataset when OFM and different single-scale atomic descriptors are taken as inputs, where the uncertainties are the standard deviations of test scores on five random test sets. The unit of the magnetic moment is $\mu_{B}$.}
	\centering
	\def\temptablewidth{1.06\textwidth}
	{\rule{\temptablewidth}{0.6pt}} 
	\begin{tabular*}{\temptablewidth}{@{\extracolsep{\fill}}cccccccccc}
		\hline
		&&OFM&NE0&NE1&NE2&NE3&NE4&NE5&\\
		\hline
		&MAE&0.860$\pm$0.107&0.997$\pm$0.086&\textbf{0.805}$\pm$0.066&0.898$\pm$0.070&0.956$\pm$0.055&1.039$\pm$0.073&0.997$\pm$0.081&\\
		\hline
		&MSE&\textbf{1.477}$\pm$0.230&1.798$\pm$0.281&1.536$\pm$0.224&1.822$\pm$0.305&2.067$\pm$0.300&2.271$\pm$0.354&2.085$\pm$0.335&\\
		\hline
		&R2&\textbf{0.684}$\pm$0.059&0.620$\pm$0.053&0.674$\pm$0.049&0.614$\pm$0.062&0.561$\pm$0.063&0.518$\pm$0.077&0.558$\pm$0.068&\\
		\hline
		&PCC&\textbf{0.836}$\pm$0.025&0.797$\pm$0.025&0.823$\pm$0.031&0.785$\pm$0.040&0.756$\pm$0.040&0.727$\pm$0.050&0.747$\pm$0.047&\\
	\end{tabular*}
	{\rule{\temptablewidth}{0.6pt}}
	\label{PP1}
\end{table*}

\begin{table*}[htbp]
	\caption{Prediction performance of the KRR model on the experimental magnetic moment dataset when different multi-scale self-supervised atomic descriptors as well as NE01+OFM are taken as inputs, where the uncertainties are the standard deviations of test scores on five random test sets. The unit of the magnetic momen is $\mu_{B}$.}
	\centering
	\def\temptablewidth{1.00\textwidth}
	{\rule{\temptablewidth}{0.6pt}} 
	\begin{tabular*}{\temptablewidth}{@{\extracolsep{\fill}}ccccccccc}
		\hline
		&&NE01&NE012&NE0123&NE01234&NE012345&NE01+OFM&\\
		\hline
		&MAE&\textbf{0.719}$\pm$0.054&0.726$\pm$0.043&0.737$\pm$0.043&0.749$\pm$0.039&0.739$\pm$0.044&0.800$\pm$0.070&\\
		\hline
		&MSE&\textbf{1.289}$\pm$0.226&1.318$\pm$0.235&1.334$\pm$0.215&1.341$\pm$0.227&1.330$\pm$0.227&1.371$\pm$0.241&\\
		\hline
		&R2&\textbf{0.727}$\pm$0.045&0.721$\pm$0.046&0.718$\pm$0.043&0.716$\pm$0.045&0.719$\pm$0.044&0.710$\pm$0.049&\\
		\hline
		&PCC&0.854$\pm$0.027&0.851$\pm$0.282&0.849$\pm$0.026&0.848$\pm$0.028&0.849$\pm$0.027&\textbf{0.855}$\pm$0.028&\\
	\end{tabular*}
	{\rule{\temptablewidth}{0.6pt}}
	\label{PP2}
\end{table*}

The prediction performance of self-supervised atomic descriptors from all the scales are shown in Supplementary Table \ref{PP1} and \ref{PP2}. From Table \ref{PP1}, we can see that the prediction ability of self-supervised atomic representations of different GNN layers varies considerably, and the overall trend of MAE from NE1 to NE5 is consistent with the cosine similarity shown in Supplementary Fig. \ref{similarity}, which indicates the significance of the distinguishability of atomic representations in property prediction. However, the initial elemental embedding NE0 is an exception. It shows the lowest cosine similarity among different representations but the highest MAE in predicting magnetic moments. The exception lies in the fact that NE0 contains no information of specific crystal environments through convolutions. It contains only some low-level information of materials through the weights of the linear layer, which is not relevant enough to magnetic moments. From NE01 to NE012345, with the information of a more extensive range, the performance is, however, declining. The reason might be that the larger-scale information is more like the environmental noises and irrelevant to the magnetic moments, which is related to the fact that the magnetic moments of transition metals and lanthanides is mainly affected by local crystal field and spin-orbit coupling, regardless of higher-order environmental corrections. 

\begin{figure}[htbp]
	\centering
	\includegraphics[height=6.0cm,width=9.0cm]{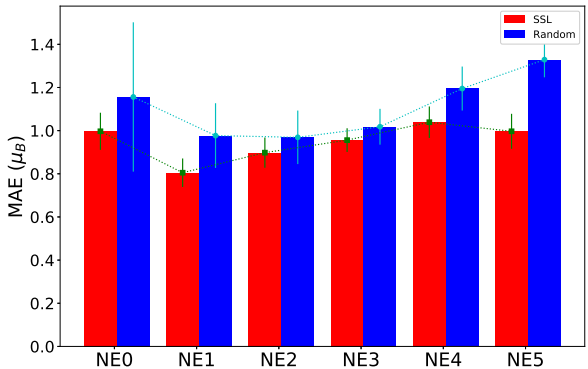}
	\caption{The MAE of the KRR model on the experimental magnetic moment dataset when the self-supervised atomic representations, SSL(red bars), and random atomic representations, Random(blue bars), are taken as inputs. The green (cyan) vertical lines denote uncertainties for SSL(Random).}
	\label{Comp}
\end{figure}

In order to highlight the benefits of self-supervised learning, we compare the atomic representations NE-SSL generated by the self-supervised pre-trained model (single-scale descriptor in Supplementary Table \ref{PP1}) and the atomic representations NE-Random generated by the GNN model with randomly initialized weights, whose MAE are shown in Supplementary Fig. \ref{Comp}. First of all, we should note that the performance of NE-Random of different GNN layers is not as bad as expected. It makes sense due to two facts: (1) the input crystal graphs already contain a rich relational inductive bias; (2) the atomic representations obtained by the message passing mechanism can receive the information about elements and structures from the crystal graphs. Importantly, we can observe that the MAE of atomic representations NE-SSL of different GNN layers is systematically lower than the MAE of NE-Random, with a much smaller standard deviation, which undoubtedly proves that the self-supervised learning can indeed optimize the generated atomic representations. 

\newpage

\section{The t-SNE plot of NE01 on local environments}

\begin{figure}[htbp]
	\centering
	\subfigure[]{
		\begin{minipage}[t]{0.4\linewidth}
			\centering
			\includegraphics[height=4.0cm,width=5.2cm]{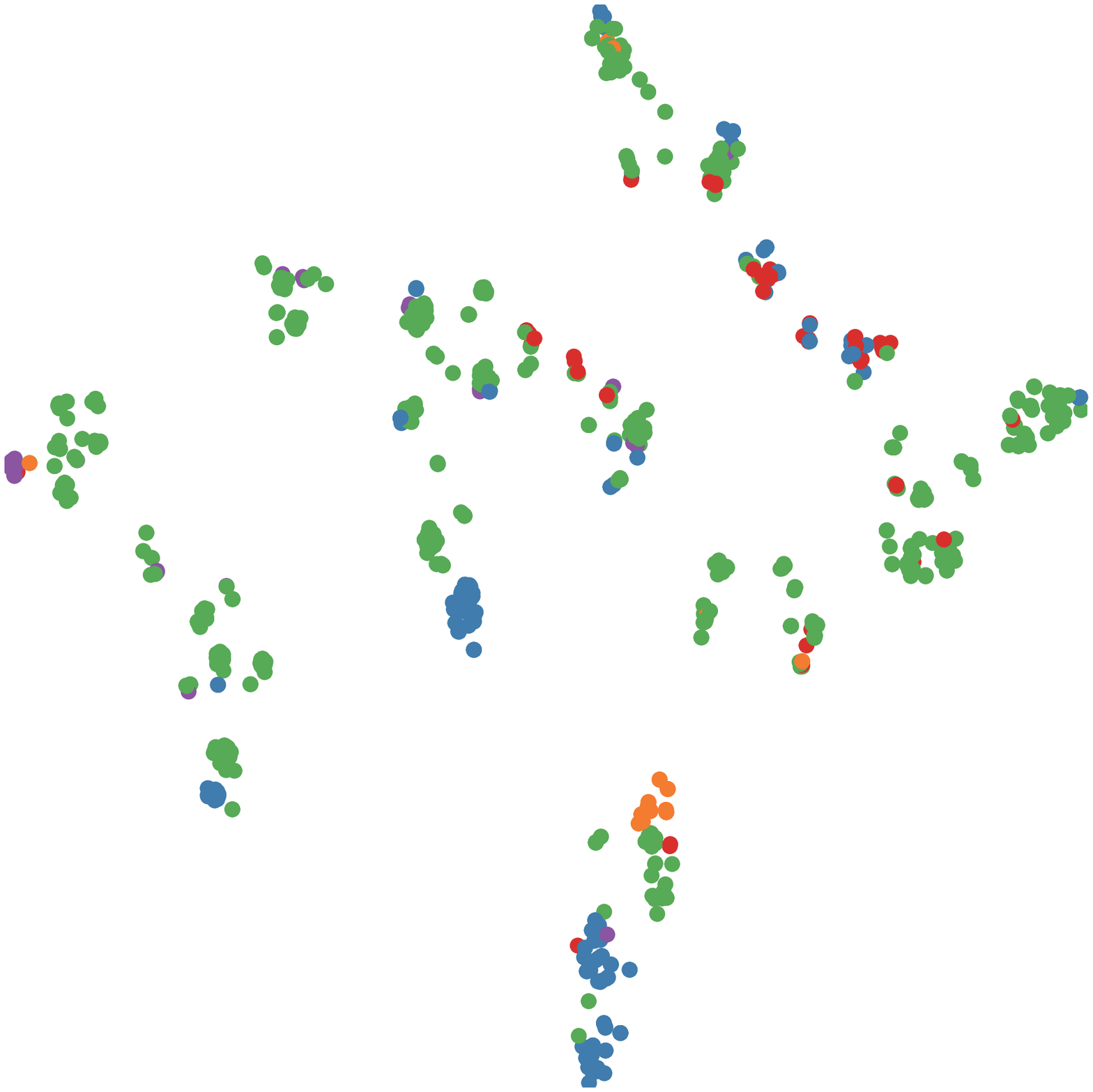}
			\label{RL}
	\end{minipage} }
	\hspace{.20in}
	\subfigure[]{
		\begin{minipage}[t]{0.4\linewidth}
			\centering
			\includegraphics[height=4.2cm,width=4.5cm]{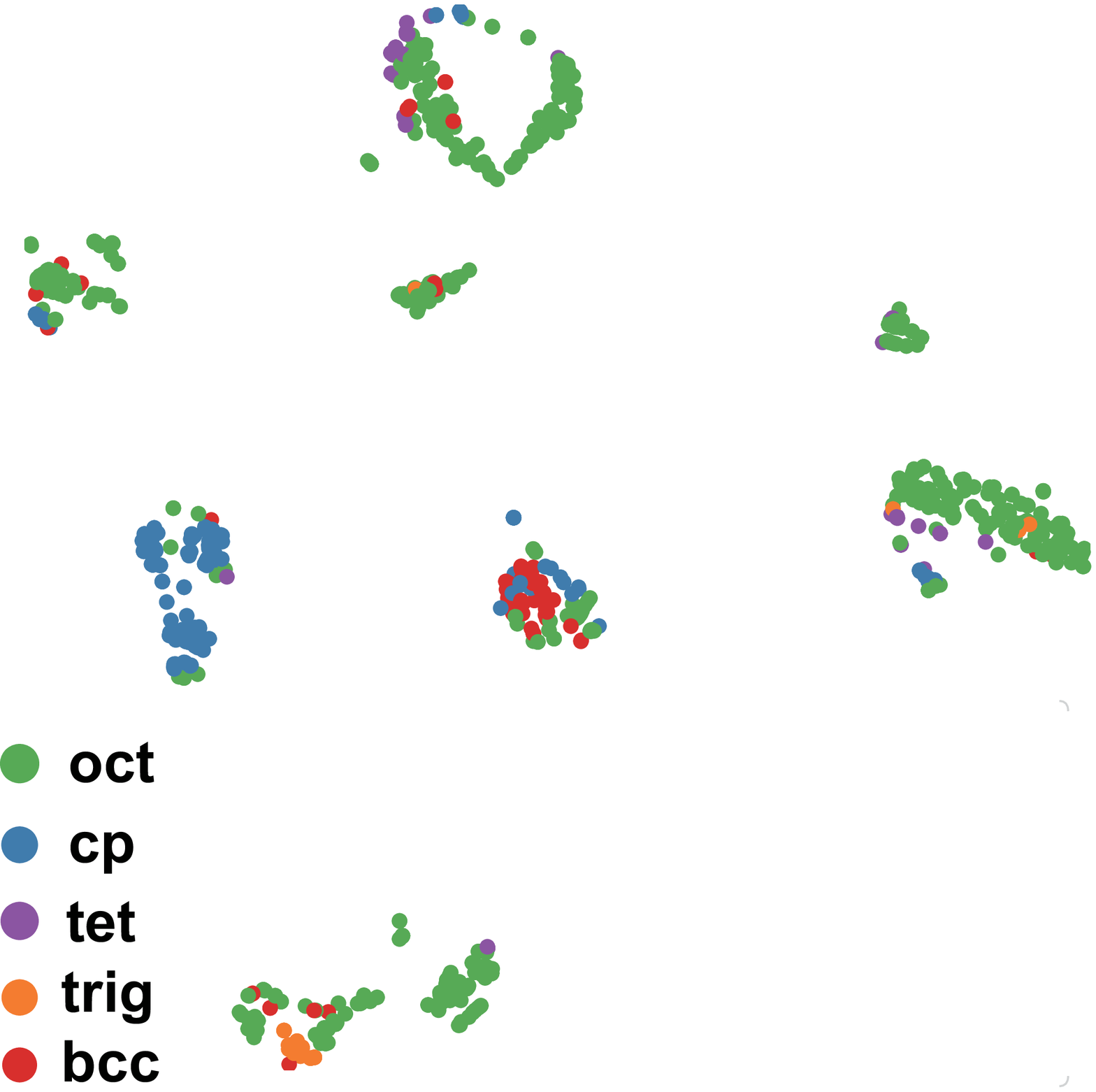}
			\label{SL1}
	\end{minipage} }
\caption{t-SNE visualization of self-supervised atomic vectors NE-SSL and random atomic vectors NE-Random with local environment labels, where the different colors represent different local environments. (a) 2D distribution of NE-Random. (b) 2D distribution of NE-SSL. The meaning of abbreviations is listed as follows: \textit{oct} means octahedral, \textit{cp} means close-packed, \textit{bcc} means body-centered cubic, \textit{tet} means tetrahedron, and \textit{trig} means trigonal bipyramidal.}
\end{figure}

We use Pymatgen's tools to identify the local environment around an atom located in the crystal. For the generated magnetic moment dataset, there are 685 magnetic atoms in the octahedral local environment, 186 atoms in the close-packed local environment, 69 atoms in the body-centered cubic local environment, 40 atoms in the tetrahedron local environment, and 35 atoms in the trigonal bipyramidal local environment. The tool cannot accurately identify the remaining 800 magnetic atoms, so they are not analyzed in this study. 

The distribution of the local environments is shown in Supplementary Figs. \ref{RL} and \ref{SL1}, where we can see that the octahedral local environment is the most typical type since magnetic atoms favor local octahedral environments to generate significant magnetic moments. The distribution of the local octahedron environment is dispersed into several small clusters due to the competition between different types of elements. We can see that the distribution of NE-SSL has more apparent patterns than of NE-Random. On the one hand, the distribution of the local octahedral environment is more concentrated with relatively large clusters; on the other hand, the patterns of body-centered cubic and local tetrahedral environments are also apparent.